\documentclass[
 aps,pra,
 amsmath,amssymb,
 reprint,
]{revtex4-2}

\usepackage{graphicx} % Include figure files
\usepackage{dcolumn} % Align table columns on decimal point
\usepackage{bm} % Bold math
\usepackage{amssymb,amsmath,amsfonts}
\usepackage{hyperref}

\usepackage[utf8]{inputenc}
\usepackage[T1]{fontenc}
\usepackage{mathptmx}
\usepackage{etoolbox}
\usepackage[figuresleft]{rotating}

\usepackage{braket}
\usepackage{qcircuit}
\hypersetup{
	colorlinks = true,
	urlcolor   = [RGB]{0,172,172},
	linkcolor  = [RGB]{0,172,172},
	citecolor  = [RGB]{0,172,172}
}

\renewcommand*{\phi}{\varphi}
\renewcommand*{\epsilon}{\varepsilon}
\DeclareMathOperator{\Tr}{Tr}

\makeatletter
\def\@email#1#2{
 \endgroup
 \patchcmd{\titleblock@produce}
  {\frontmatter@RRAPformat}
  {\frontmatter@RRAPformat{\produce@RRAP{*#1\href{mailto:#2}{#2}}}\frontmatter@RRAPformat}
  {}{}
}
\makeatother

\begin{document}

\preprint{AIP/123-QED}

\title[Effective programming of a photonic processor with complex interferometric structure]{Effective programming of a photonic processor with complex interferometric structure
}

% \author{\newline}
\author{I.V.~Kondratyev$^{1}$}
\email{iv.kondratjev@physics.msu.ru}
\author{K.N.~Urusova$^{1}$}
\author{A.S.~Argenchiev$^{1}$}
\author{N.S.~Klushnikov$^{1}$}
\author{S.S.~Kuzmin$^{1,2}$}
\author{N.N.~Skryabin$^{1,2}$}
% \author{\newline}
% \author{\newline}
\author{\\A.D.~Golikov$^{3}$}
\author{V.V.~Kovalyuk$^{3,4}$}
\author{G.N.~Goltsman$^{2,3}$}
\author{I.V.~Dyakonov$^{1,2}$}
\author{S.S.~Straupe$^{1,2}$}
\author{S.P.~Kulik$^{1}$}

\affiliation{$^1$Quantum Technology Centre and Faculty of Physics, M.\,V. Lomonosov Moscow State University, 1 Leninskie Gory, Moscow, 119991, Russia}
\affiliation{$^2$Russian Quantum Center, 30 Bolshoy Boulevard, building 1, Moscow, 121205, Russia}
\affiliation{$^3$Department of Physics, Moscow State Pedagogical University, Moscow 119992, Russia}
\affiliation{$^4$Laboratory of Photonic Gas Sensors, University of Science and Technology MISIS, Moscow 119049, Russia}

\date{\today}

\begin{abstract}
Reconfigurable photonics have rapidly become an invaluable tool for information processing. Light-based computing accelerators are promising for boosting neural network learning and inference \cite{xiao2021large, meng2023compact, cheng2024multimodal}, and optical interconnects are foreseen as a solution to the information transfer bottleneck in high-performance computing \cite{li2021scaling, rizzo2022petabit, netherton2024high}. In this study, we demonstrate the successful programming of a transformation implemented using a reconfigurable photonic circuit with a non-conventional architecture. The core of most photonic processors is an MZI-based architecture \cite{Reck1994, Clements16} that establishes an analytical connection between controllable parameters and circuit transformation. However, several architectures \cite{tang2017integrated, Robust2020} that are substantially more difficult to program have improved robustness to fabrication defects. We use two algorithms that rely on different initial datasets to reconstruct the circuit model of a complex interferometer, and then program the required unitary transformation. Both methods performed accurate circuit programming with an average fidelity greater than 98\%. Our results provide a strong foundation for the introduction of non-conventional interferometric architectures for photonic information processing.
\end{abstract}

\maketitle

\section{\label{sec:introduction} Introduction}

Integrated photonics has significantly affected communications \cite{Tanizawa15, Lu16, Suzuki17}, quantum and classical computing \cite{Capmany2020}, optical neural networks \cite{zhang2021optical, meng2023compact, cheng2024multimodal}, and life sciences \cite{goel2020survey, ciceri2025high}. Its core capability—on‑chip phase modulation—enables the routing and manipulation of optical states, allowing interferometric waveguide circuits to implement programmable information processing algorithms in light. These systems offer a wide operational bandwidth and superior energy efficiency per operation \cite{shekhar2024roadmapping,lin2024power, butt2025lighting}.

In this study, we focus on photonic device programming for information processing. A typical photonic chip for this purpose includes multiple power splitters and phase shifters comprising a large multiarm interferometer \cite{Carolan2015,taballione2021universal, barzaghi2025low}. Once a photonic processor has been fabricated, it must be properly calibrated to implement a unitary transformation on demand. This process is challenging if the architecture of the interferometric circuit includes complex elements, such as multiport power splitting circuits \cite{kondratyev2020multiport, meng2023compact, cheng2024multimodal}.

The transformation of the interferometer is described by a unitary transfer matrix $U$, which depends on the phases between different paths $U = U(\vec{\varphi})$. Typically, the procedure for calibrating phase shifters involves injecting coherent light and control signal through the phase shifters one by one and recording the output interference pattern. The data collected are fitted with an interferometer model accompanied by phase functions $\phi(X)$, where $X$ is a control signal (typically current $I$ or voltage $V$). The result of the procedure are the exact functions $\phi(X)$ that characterize each phase shifter. 

\begin{figure*}[ht!]
\centering
\includegraphics[width=1\linewidth]{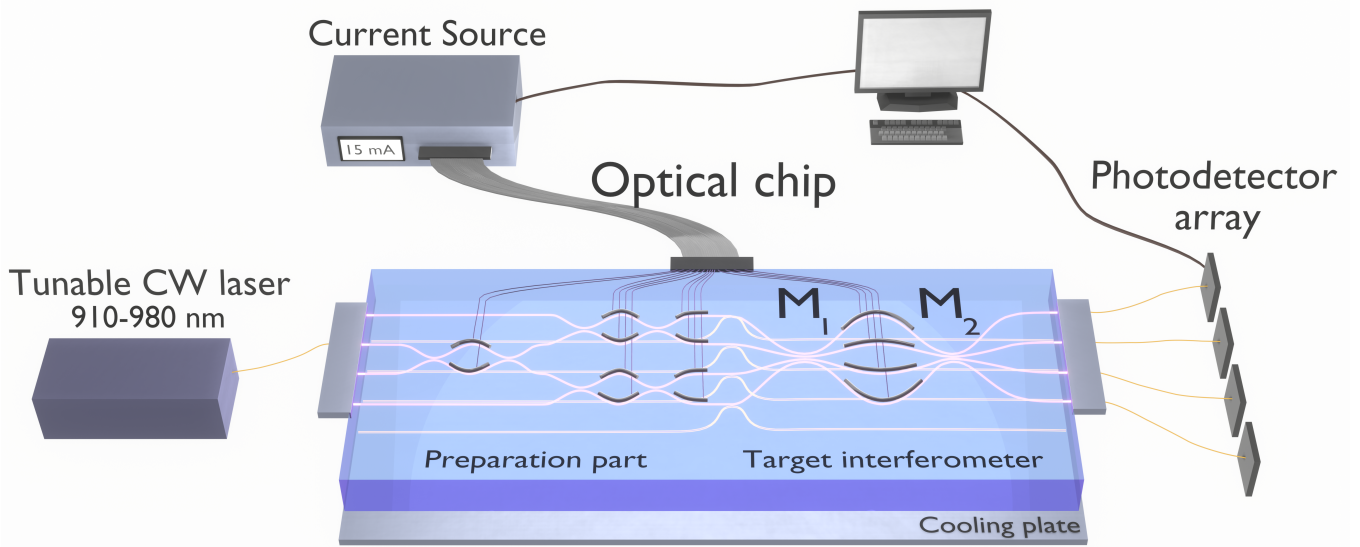}
\caption{Schematic photonic chip structure. The preparation part, consisting of three cascaded MZIs and a target complex interferometer consisting of two $4\times 4$ multiport couplers sandwiching three tunable phase shifters, are both in the same fused silica sample. Laser light is injected into the second input port of the chip. This allows for the preparation of an arbitrary optical distribution of light among the four input ports of the target interferometer. 
The MZIs in the interferometer can be individually calibrated using auxiliary waveguides that are weakly coupled to the output section of the preparation part of the interferometer. Auxiliary waveguides lie 35~$\mu$m below the main waveguide plane with preparation and target interferometers. Each auxiliary waveguide is only coupled to the corresponding waveguide of the preparation interferometer and only at its output region.} 
\label{fig:experimental setup}
\end{figure*}

Common photonic architectures rely on a Mach-Zehnder interferometer (MZI) as a building block \cite{Reck1994, Clements16, mojaver2023addressing, de2025new}. This class of architectures is based on the well-known isomorphism of the MZI transfer matrix and arbitrary unitary rotation $SU(2)$. Within MZI-based architectures, calibration can be performed for each interferometer one after another, as it is possible to isolate the effect of each individual block \cite{Carolan2015, zhang2021optical, skryabin2023two, taballione202320, pentangelo2024high, lin2024power}. 

However, this strategy is not applicable to interferometers with other types of building blocks. For example, using multiport beamsplitters \cite{Robust2020} or beamsplitter meshes \cite{Fldzhyan20} turns out to be beneficial in minimizing the effects of fabrication imperfections. Interferometers of this kind have already been demonstrated \cite{tang2017integrated, zhou2018tunable, kondratyev2024large}, but programming a desired unitary transformation remains an unresolved challenge. The well-known issue of the crosstalk between multiple tunable phase shifters complicates the unitary programming problem even more by rendering easy calibration procedures inaccurate even in basic architectures such as \cite{Reck1994, Clements16}.

The task of programming a unitary transfer matrix of a reconfigurable photonic system can be approached differently. The first step would be to infer the parameters of the photonic system and construct its digital replica. The parameters include both characteristics of optical components, such as, for example, the power splitting ratio of the beamsplitters, and the coefficients connecting tunable phase shifts with applied control signal. Several recent studies followed these guidelines \cite{bandyopadhyay2022single, pentangelo2024high, fyrillas2024scalable, barzaghi2025low, fyrillas2025resource} and demonstrated programming transfer matrices of MZI mesh-based interferometric systems. 

However, there exists substantial practical interest in interferometers with $n \times n$ waveguide splitters \cite{tang2021ten, xiao2021unitary, yang2024programmable}, as they can be used as a powerful tool for photonic convolutional networks and random matrix generation \cite{meng2023compact, zelaya2025integrated}. %We will call the interferometers consisting of such couplers multiport interferometers. 
Meanwhile, there is no straightforward approach to enable the programming of unitary transfer matrices of the interferometer with $n\times n$ splitters. 

It should be mentioned that many of the existing studies on MZI-based interferometer programming have not used the actual matrix fidelity measure to demonstrate the ability to achieve the desired unitary matrix set on a chip. Instead, in these studies, another metric called amplitude fidelity was used \cite{taballione202320, lin2024power, fyrillas2024scalable, barzaghi2025low, fyrillas2025resource}, which is inherently insensitive to the angles of the unitary transformation that are crucial for the evolution of multiphoton states through linear optical interferometers. However, amplitude fidelity has been argued to contain nearly the same amount of information, demonstrating the universal capabilities of the interferometers under study \cite{pentangelo2024high}.

Several recent theoretical proposals suggested ways to program an arbitrary complex interferometer \cite{kuzmin2021architecture,bantysh2023fast}. In these proposals, the digital model of the interferometer is inferred from the experimental data set and then serves as a reliable tool to determine the phase configuration corresponding to the required unitary transfer matrix. Both methods \cite{kuzmin2021architecture,bantysh2023fast} require precise control of phase shifts $P(\vec{\varphi})$ which, in turn, demand preliminary fine calibration $\vec{\varphi} = \vec{\varphi} (\vec{X})$ of the phase modulators.

Herein, we present the programming algorithm based on the reconstructed digital model of the photonic interferometer with arbitrary architecture. We used the algorithm to demonstrate the reconfigurable operation of the 4-mode multiport photonic processor. To verify the algorithm, we experimentally set and measured 100 random \textit{unitary} transformations on the chip and compared them with those predicted using the chips' digital model. We used the true matrix fidelity measure for comparison and reported average fidelity $99.6\pm 0.2$.

\begin{figure*}[ht]
\centering
\includegraphics[width=1\linewidth]{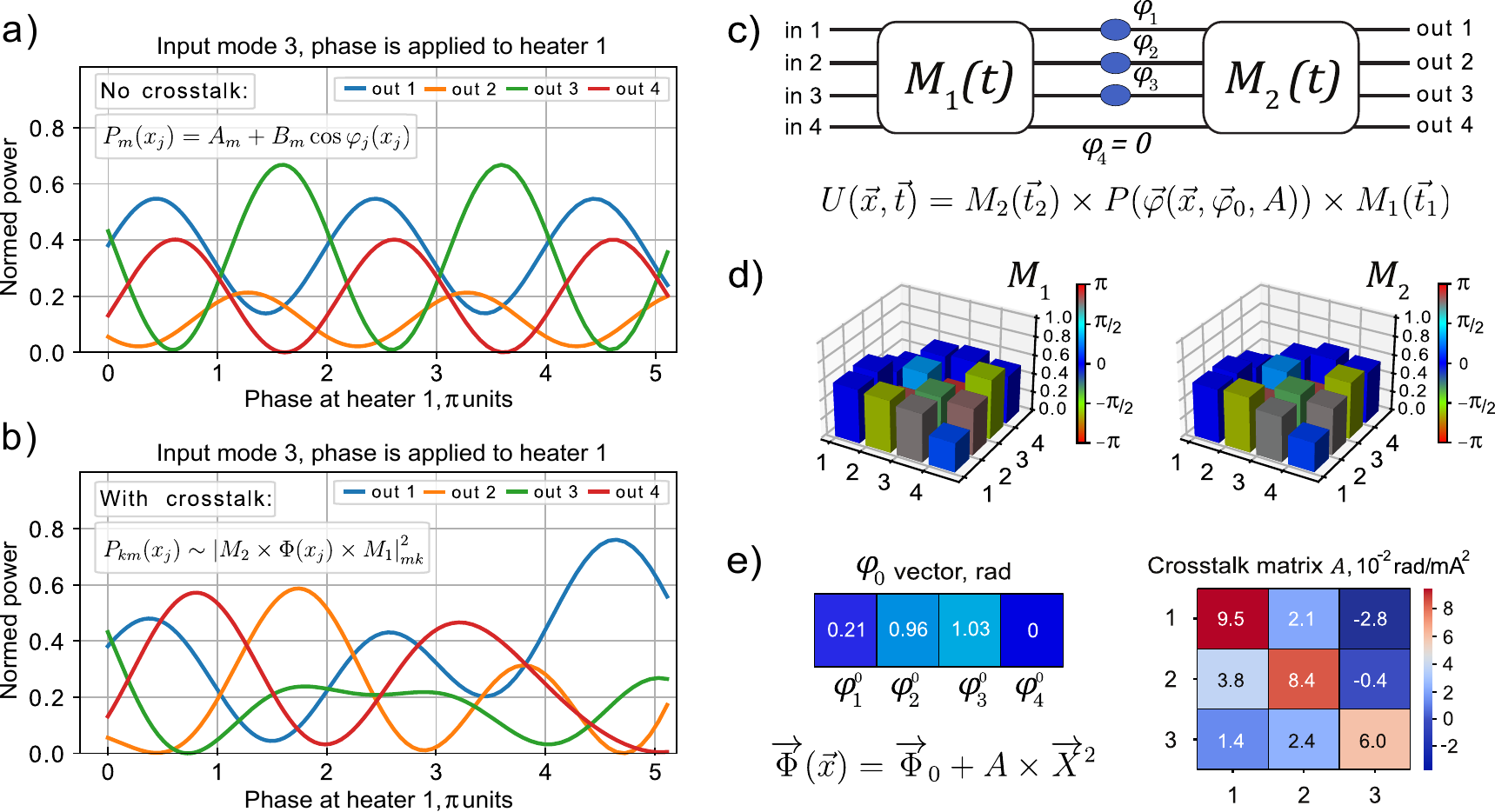}
\caption{ a) Typical output optical power dependence $P_m(\varphi(x_j))$ in cases a) without crosstalk and b) with mutual crosstalk between adjacent phase-shifters. This illustration shows the numerical modeling of the calibration of the phase shifter $\varphi_1$ when coherent radiation is injected into the third input port of the target interferometer (see Fig. \ref{fig:experimental setup}). In the absence of crosstalk, all $P_m(\varphi(x_j))$ curves are sinusoidal with an equal period. However, in the presence of crosstalk, each $P_m(\varphi(x_j))$ becomes a complex and non-periodic curve. c) A schematic illustration of the digital model of the target interferometer and its parametrization. The chips' unitary transformation $U$ was decomposed in three blocks: two mode-mixing blocks $M_{1,2}$ and a phase shift layer $P(\vec{\varphi}(\vec{x}, \vec{\varphi}_0, A))$ between them. Both mode-mixing blocks $M_{1,2}$ were parameterized by a triangular mesh of MZIs \cite{Reck1994}, which required 9 real parameters $t_j$ in the range $t_j \in [ 0, 2 \pi ]$. d) Mode-mixing unitary matrices $M_{1,2}$ obtained from the simultaneous approximation of all calibration data.  e) The $\Phi_0$ vector of the bias phase shifts and the matrix of crosstalk $A$ obtained from the simultaneous approximation of the calibration data.}  
\label{fig:with_or_withot_crosstalk}
\end{figure*}

\section{ Methods }
\subsection{ Optical Chip Design }

We demonstrate our programming algorithm on a photonic chip that was manufactured using femtosecond laser writing (FSLW) in a fused silica glass sample \cite{Cai22}. The description of the waveguide fabrication regime can be found in the Appendix \ref{app:fab_process}. The single-mode waveguide structure of the chip contained a target $4\times 4$ multiport interferometer and a preparation interferometer. The preparation interferometer is a mesh of three MZIs which can set an arbitrary amplitude distribution at the output of the preparation part. We use the preparation interferometer as a tool for calibrating the target interferometer and probing its transfer matrix. Both the preparation and the target interferometers are equipped with tunable thermo-optical phase shifters, which are also fabricated using the femtosecond laser writing setup.

The FSLW technology allows us to create waveguides inside a three-dimensional space. This feature enables us to write four auxiliary waveguides $35$~$\mu$m below the main waveguide plane. These waveguides are evanescently coupled to the output section of each waveguide of the preparation interferometer. We used these waveguides to monitor the amplitudes fed to the input of the target interferometer, which is a crucial requirement for reconstruction of the digital model of the target interferometer. 

The light from the chips' output ports was collected using a fiber array, and the corresponding optical power was then measured using four photodetectors. The input fiber and the output fiber array were mounted on high-precision six-axis positioners. We could use the output positioner to switch between the output of the target interferometer and the auxiliary waveguides. The structure of the chip and the experimental setup are shown schematically in Fig. \ref{fig:experimental setup}.

Calibration of the MZIs-based interferometer is a known procedure that is performed by individually calibrating each of its constituent MZIs \cite{Carolan2015, skryabin2023two, taballione202320, lin2024power}. We used the same principles to calibrate the preparation interferometer. More details on the optical chip design and preparation for interferometer calibration can be found in Appendices \ref{app:chip_design} and \ref{app:prep_MZI_calibration}.

The target interferometer consists of two $4\times 4$ multiport couplers sandwiching three tunable phase shifters. The optical interferometers with this kind of architecture were studied in \cite{kondratyev2020multiport}. The unitary transformation of the target interferometer can be expressed as
\begin{equation}\label{eq::unitary_chip_matrix}
    U(\vec{\varphi}) = M_2 \times P(\vec{\varphi}) \times M_1,
\end{equation}
where $M_{1,2}$ are the unitary matrices corresponding to each of the $4\times 4$ multiport couplers, which we also refer to as constant mode mixing blocks, $\vec{\varphi} = \{\varphi_1, \varphi_2, \varphi_3 \}$ is a vector of three phase shifts and $P(\vec{\varphi}) = diag(e^{i\varphi_1},e^{i\varphi_2},e^{i\varphi_3}, 1)$ is a diagonal matrix of the phaseshifting layer. The key task for successful implementation of the programming procedure is the reconstruction of the two matrices $M_{1,2}$ and $P(\vec{\varphi}(\vec{X}))$, where $\vec{X} = \{ x_1, x_2, x_3 \}$ are the electrical currents applied to thermo-optical phase shifters.

\subsection{ Target interferometer calibration }

The calibration of tunable phase shifters in an $N$-port reconfigurable interferometer starts with injecting coherent laser light into the sample. For each input port, the electrical current $x_j$ through the $j$-th phase shifter is swept within a range of values while measuring the optical power $P_m$ from each output port $m$ of the interferometer $P_m = P_m(x_j)$.

\begin{figure}[ht!]
\centering
\includegraphics[width=1\linewidth]{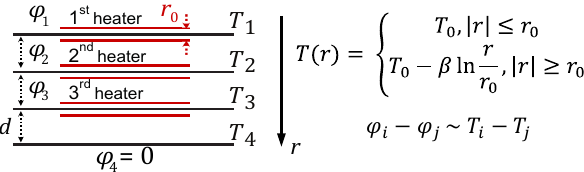}
\caption{ Guidelines for the  estimation of relations between the crosstalk elements $A = \{ \alpha_{ij} \}$. } 
\label{fig:crosstalk_estim}
\end{figure}

The ideal scenario occurs when $j$-th modulator affects only the corresponding shift on the interferometer $\varphi_j = \varphi(x_j)$. In this case, the output power $P_m$ obeys the periodic law
\begin{equation}\label{eq::power_no_crosstalk}
    P_m(x_j) = A_m + B_m \cos{\varphi_j(x_j)},
\end{equation}
where $A_m$ and $B_m$ are real constants (independent of $x_j$) depending on the inner interferometer structure, in our case the matrices $M_{1,2}$ describing the multiport splitters. The $\varphi = \varphi_j(x_j)$ is a specific dependence of the $j$-th phase shift on the corresponding driving signal $x_j$ and is generally assumed to be a polynomial function of $x_j$. For example, in the case of a thermo-optical phase shifter, $\varphi_j(x_j) = \varphi^{0}_j + \alpha_j x_j^2$ because the induced phase shift is directly proportional to the thermal power dissipated by a resistive electrode. A typical $P_m(\varphi(x_j))$ behavior in the absence of crosstalk is shown in Fig. \ref{fig:with_or_withot_crosstalk}a.
% Of course,  $A_m \geq |B_m|$ to ensure $P_m(x_j) \geq 0$.

After measuring $P_m(x_j)$, the approximation of the experimental data with (\ref{eq::power_no_crosstalk}) provides the calibration parameters $\varphi^0_j$ and $\alpha_j$ for a specific tunable phase-shifter. Note that the specific values of the constants $A_m$ and $B_m$ are not important in this procedure, and the most critical parameters for each phase-shifter are the corresponding $\varphi^0_j$ and $\alpha_j$. After obtaining all calibration parameters $\varphi^0_j$ and $\alpha_j$ for each $j$-th tunable phase shifter, the algorithms \cite{kuzmin2021architecture, bantysh2023fast} can be implemented to reconstruct $M_{1,2}$. Once the $M_{1,2}$ are inferred, then a full digital model of the photonic chip is available to find a phase configuration corresponding to a required unitary transfer matrix.

However, in most real optical chips, even those fabricated using the most advanced technology \cite{taballione202320}, there is often an effect of mutual cross-influence between different phase shifters. This can occur because of thermal or electrical crosstalk. Due to the crosstalk effect, the optical power $P_m$ of the $m$-th output port of the chip can no longer be described by \ref{eq::power_no_crosstalk} because the approximation constants $A_m$ and $B_m$ are functions of $x_j$, as well as of the indices of the input and output ports of the interferometer. This leads to a complex and non-periodic dependence $P_m(x_j)$. Figure \ref{fig:with_or_withot_crosstalk}b shows the typical dependency $P_m(x_j)$ in the presence of crosstalk.

In our optical chip, we used thermo-optical phase shifters for which a significant crosstalk effect between adjacent heaters was observed. Due to the presence of mutual coupling between different phase shifters, the phase shift $\varphi_i$ induced in the $i$-th heater depends not only on the corresponding current $x_i$ but also on all currents $x_j$ applied to other heaters (which may influence the $i$-th phase shift):
\begin{equation}\label{eq::phi_with_crosstalk}
    \varphi_i(\vec{x}) = \varphi^{(i)}_0 + \sum_j \alpha_{ij} x_j^2,
\end{equation}
where $x_j$ is the current applied to the $j$-th heater, $\varphi^{(i)}_0$ is a constant bias phase shift that exists in the particular interferometer arm, and $\alpha_{ij}$ is a constant that determines the strength of the effect of the $j$-th heater on the $i$-th phase shifter. The relation between the phases and the currents flowing through the heaters can be rewritten in matrix form as follows:
\begin{equation}\label{eq::phi_with_crosstalk_matrix}
    \overrightarrow{\Phi}(\vec{x}) = \overrightarrow{\Phi}_0 + A \times \overrightarrow{X}^2,
\end{equation}
where $\overrightarrow{\Phi} = \{\varphi_1, \varphi_2, \varphi_3 \}^T$  is a column vector of the phase shifts to be set, $\overrightarrow{\Phi_0} = \{\varphi^{0}_1, \varphi^{0}_2, \varphi^{0}_3 \}^T$ is the column vector of the bias phase shifts, $A = \{ \alpha_{ij} \}$ is a crosstalk matrix containing all nine crosstalk constants $\alpha_{ij}$, $i,j = \overline{1,3}$, and  $\overrightarrow{X}^2 = \{x^{2}_1, x^{2}_2, x^{2}_3 \}^T$ is a column vector of squared current values.

Therefore, according to (\ref{eq::phi_with_crosstalk}), if only one $x_j$ is varied, it will affect all three phase shifts $\varphi_i$ simultaneously, leading to a complex and nonperiodic dependence of the output optical power on the applied current $x_j$ (see Fig. \ref{fig:with_or_withot_crosstalk}b). Nevertheless, if coherent light is injected into the $k$-th input port of the photonic chip, the output optical power at the $m$-th output port can be described in terms of the $m,k$-th element of the unitary matrix of the interferometer:
\begin{equation}\label{eq::power_with_crosstalk}
    P_{km}(x_j) \sim \left|  M_2 \times \Phi(x_j) \times M_1 \right|^2_{mk},
\end{equation}
which can be used to fit the corresponding experimental data, with both $M_1$, $M_2$, and $\overrightarrow{\Phi_0}$, and the matrix of crosstalk constants $A$ as optimization parameters.

Moreover, any experimentally measured optical power $P_{km}$ from the applied current for any input $k$ and output $m$ ports of the interferometer and for any heater $j$ must obey (\ref{eq::power_with_crosstalk}) with the same interferometer parameters $M_{1,2}$, $\overrightarrow{\Phi_0}$ and $A$, as they are all generated by the same photonic chip, described by its unitary matrix $U$. Once all possible optical powers from the applied current dependencies, $P_{km}(x_j)$, have been experimentally measured for each $k$, $m$, and $j$, they can be simultaneously approximated using a single interferometer model, determining the constituents of the optical chips $M_{1,2}$ and the necessary calibration relations between all currents and phase shifts $\overrightarrow{\Phi_0}$ and $A$. The key points of the approximation over the whole dataset are given in the next subsection.

\subsection{The target interferometer model}\label{sub_sec::sim_opt_details}

The approximation of the whole data set $P_{km}(x_j)$ in the form of (\ref{eq::power_with_crosstalk}) requires appropriate parameterization of the optical chip transfer matrix $U(\vec{x})$ as a function of the applied currents: 
\begin{equation}\label{eq::unitary_chip_matrix}
    U(\vec{x}, \vec{t} ) = M_2(\vec{t}_{2}) \times P(\vec{\varphi}(\vec{x}, \vec{\varphi}_0, A)) \times M_1(\vec{t}_{1}),
\end{equation}
where $\vec{t}_{1,2} = \{ t_1, \dots, t_{N} \}$ are two sets of real parameters that determine $M_{1,2}$, $P(\vec{\varphi}(\vec{x}, \vec{\varphi}_0, A)) = diag(e^{i\varphi_1}, e^{i\varphi_2}, e^{i\varphi_3}, 1)$ is a diagonal matrix of phase shifts (see Fig. \ref{fig:with_or_withot_crosstalk}c). In turn, phase shifts depend on the currents $x_j$ and the calibration parameters $\varphi^{(0)}_j$ and $\alpha_{ij}$:

\begin{equation}\label{eq::phi_with_crosstalk_matrix}
    \begin{pmatrix}
     \varphi_1\\
     \varphi_2\\
     \varphi_3\\
    \end{pmatrix}
    =
    \begin{pmatrix}
     \varphi^{(0)}_1\\
     \varphi^{(0)}_2\\
     \varphi^{(0)}_3\\
    \end{pmatrix}
    +
    \begin{pmatrix}
    \alpha_{11}, \alpha_{12}, \alpha_{13} \\
    \alpha_{21}, \alpha_{22}, \alpha_{23} \\
    \alpha_{31}, \alpha_{32}, \alpha_{33} \\
    \end{pmatrix}
    \times
    \begin{pmatrix}
     x_1^2\\
     x_2^2\\
     x_3^2\\
    \end{pmatrix},
\end{equation}
where later two are to be experimentally determined. 

\begin{figure*}[ht!]
\centering
\includegraphics[width=1\linewidth]{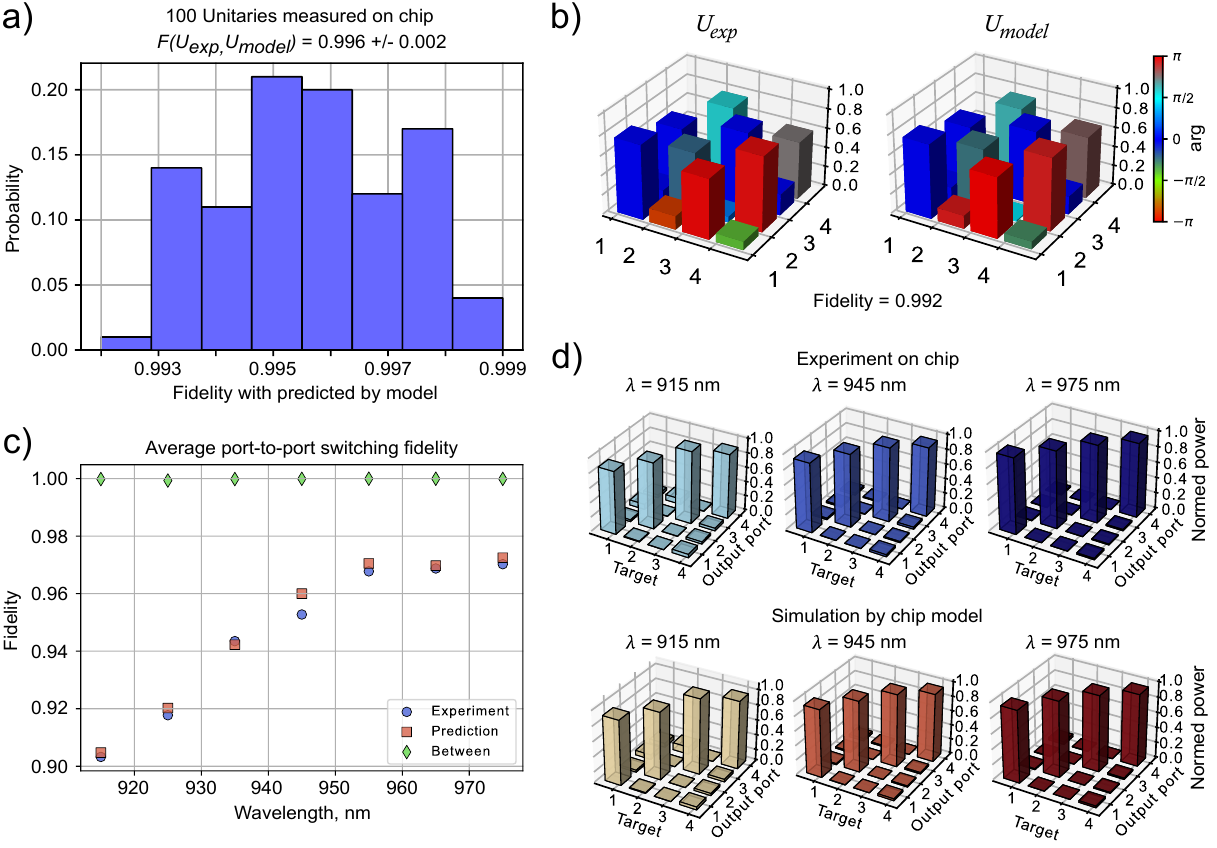}
\caption{ The results of the digital model of the optical chip quality testing. a) The Matrix fidelity between 100 measured \textit{unitaries} on chip and the corresponding unitaries simulated using the chip's model.  b) An example of a pair of measured-simulated unitaries with the lowest value of matrix fidelity $99.2\%$. All the other 99 pairs of measured-simulated unitaries have greater matrix-fidelity values. c) Graph of average optical port-to-port switching fidelities for different wavelengths of coherent radiation. d) Histograms of the power distribution at the output ports of the optical chip, reconfigured for one-to-one switching to a specific output, are shown for the first input port and three wavelengths.} 
\label{fig:main_res_100_unittaries}
\end{figure*}

For mode-mixing blocks $M_{1,2}$ we used the well-known unitary parameterization \cite{Reck1994} that for the $4 \times 4$ port interferometer requires nine real parameters varying from $0$ to $2\pi$ each (see Appendix \ref{app:global_fit_details} for details). The total number of real parameters for our photonic chip then sums up to $30$, including $9 \times 2 = 18$ parameters for $M_{1,2}$,  $3$ bias phases $\varphi^{(0)}_j$, and $9$ crosstalk constants $\alpha_{ij}$.

An adequate initial guess for the optimization parameters is important because the global optimization process experiences difficulties converging to the global minimum using a random initial seed. The specific physical properties of the experimental system should be considered when setting the limits of the parameter values.

\section{Experimental results}\label{sec::results}

We gather the data set $P_{km}(x_j)$ and run the approximation which yielded $M_{1}$ and $M_{2}$, $\overrightarrow{\Phi_0}$ and $A$, which are illustrated in Fig. \ref{fig:with_or_withot_crosstalk}d,e. Although no additional constraints were set for parameterization of $M_{1}$ and $M_2$, they appeared to be almost identical $M_1 \approx M_2$ as shown in Fig. \ref{fig:with_or_withot_crosstalk}d. This fact comes from the stability of our FSLW setup. Moreover, from Fig. \ref{fig:with_or_withot_crosstalk}e we can see that the bias phases satisfy $\varphi^0_1 \approx \varphi^0_4$ and $\varphi^0_2 \approx \varphi^0_3$, which can also be explained by the actual geometry of the phase shift layer $P(\varphi)$ on our chip (see Fig. \ref{fig:experimental setup}). In the phase-shift layer of the target interferometer, the first and fourth waveguides have the same length, while the second and third waveguides also have the same length but are shorter than the first and fourth. This results in non-zero bias phase shifts, as shown in Fig. \ref{fig:with_or_withot_crosstalk}e.

The reconstructed elements of the crosstalk matrix $A$ (see Fig. \ref{fig:with_or_withot_crosstalk}e) correlate with our assumption about their mutual relation based on the geometry of the optical structure. The negative values of some of the elements are also predicted by our simple model. Qualitative estimation of relative magnitudes of $a_{ij}$ can be made using the linearity of the heat equation, the assumption of logarithmic temperature decay with distance from the heater and the proportionality between phase and temperature differences $\varphi_i - \varphi_j \sim T_i - T_j$. Assuming (endless) parallel straight heaters with a width of $r_0$ right above (endless) parallel straight waveguides are equally separated by $d$ from each other (see Fig. \ref{fig:crosstalk_estim}), one can derive the following form of the crosstalk matrix $A$:

\begin{equation}\label{eq::crosstalk_relation}
    \centering
    A = \{ \alpha_{ij} \} =  \alpha_0
    \begin{pmatrix}
    \ln{ \frac{3d}{r_0}} & \ln{2} & -\ln{2} \\
    \ln{3} & \ln{ \frac{2d}{r_0}} & 0 \\
    \ln{ \frac{3}{2} } & \ln{ 2 } & \ln{ \frac{d}{r_0} } \\
    \end{pmatrix},
\end{equation}
where $\alpha_0$ is a constant (see Appendix \ref{app:global_fit_details} for more details).

% \begin{figure*}[ht!]
% \centering
% \includegraphics[width=1\linewidth]{Figs/Main_Results_edit.pdf}
% \caption{ The results of the digital model of the optical chip quality testing. a) The Matrix fidelity between 100 measured \textit{unitaries} on chip and the corresponding unitaries simulated using the chip's model.  b) An example of a pair of measured-simulated unitaries with the lowest value of matrix fidelity $99.2\%$. All the other 99 pairs of measured-simulated unitaries have greater matrix-fidelity values. c) Graph of average optical port-to-port switching fidelities for different wavelengths of coherent radiation. d) Histograms of the power distribution at the output ports of the optical chip, reconfigured for one-to-one switching to a specific output, are shown for the first input port and three wavelengths.} 
% \label{fig:main_res_100_unittaries}
% \end{figure*}

\subsection{Testing the digital model of the chip}\label{sub_sec::model_test}
To verify the validity of our digital interferometer model, we performed the following procedure. We sampled a set of 100 random current configurations $\{ x^{(n)}_1, x^{(n)}_2, x^{(n)}_3 \}$ ($n = \overline{1,100}$) in the range of 0 to 15 mA and applied them to our photonic chip. For each set of currents, the \textit{unitary} transformation of the chip $U^{(n)}_{exp}$ was measured directly using coherent CW radiation \cite{rahimi2013direct,heilmann2015novel} (see Appendix \ref{app:unitaries_measure} for details). On the other hand, $U^{(n)}$ can be simulated from the digital chip model (\ref{eq::unitary_chip_matrix}) and compared to the one measured by the matrix fidelity measure:
\begin{equation}\label{eq::matrix_fidelity_form}
    F(U_{exp}, U_{sim}) = \frac{ \Tr(U_{exp} \times U_{sim}) \Tr({U_{sim} \times U_{exp}})}{ \Tr(U_{exp} \times U_{exp}) \Tr(U_{sim} \times U_{sim}) },
\end{equation}

The results are shown in Fig. \ref{fig:main_res_100_unittaries}a. The average fidelity of 100 measured \textit{unitary} matrices is $99.6 \pm 0.2 \%$. Figure \ref{fig:main_res_100_unittaries}b shows the measured unitary matrices and the chip model unitary matrices for specific currents on a chip, with the lowest mutual matrix fidelity among all sets. Remarkably, the lowest fidelity between the measured and simulated unitaries was $99.2\%$. This result clearly demonstrates the high quality of the digital model of our photonic chip. 
%and demonstrates the robustness of the multiport interferometer architecture to the fabrication defects which result in the deviation of the mode-mixing blocks  $M_{1}, M_{2}$ from the Fourier matrix.

\begin{figure*}[t!]
\centering
\includegraphics[width=0.9\linewidth]{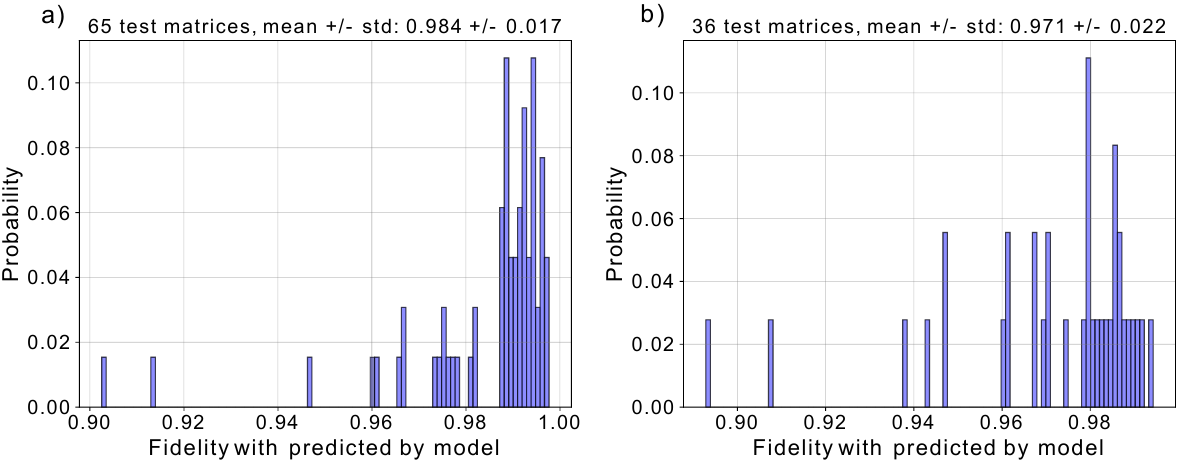}
\caption{ The fidelity distribution histograms for ML based model test. a) Testing the learning of artificial neural network on phase shifts-matrices ($\vec{\varphi}$-$U$) dataset. b)  Testing the learning of artificial neural network on currents-matrices ($\vec{x}$ - $U$) dataset. } 
\label{fig:ML_histograms}
\end{figure*}

\subsection{Machine learning based method for digital model reconstruction}

We performed a proof-of-concept experiment to reconstruct the digital interferometer model and perform transfer matrix programming according to \cite{kuzmin2021architecture}, which is based on the different data sets.

This method uses the training and validation data sets comprised of pre-reconstructed unitaries corresponding to the values of the driving signals $\vec{x}$ (currents through the heaters in our case) or the phase shift values $\vec{\phi}$ induced by the driving signals. We used the calibration $\phi(x)$ and the unitary matrices that were previously measured (see Fig. \ref{fig:main_res_100_unittaries}a) and prepared two data sets. The first data set $(\vec{\phi},U)$ includes the phase shifts $\vec{\phi}$ and the corresponding matrices $U$ and the second data set $(\vec{x},U)$ includes the current values $\vec{x}$ and the corresponding matrices $U$. We then used both data sets to estimate the performance of the method and compare it with the results provided by the global optimization. The details of the learning process can be found in Appendices \ref{app:ml-based-dataset-phases} and \ref{app:ml-based-dataset-currents}.

The method \cite{kuzmin2021architecture} infers the digital interferometer model
\begin{equation}
    U = A_2(\vec{p}_2) \times P \times A_1(\vec{p}_1),
\end{equation}
where $P$ is a diagonal matrix of phase shifts and $A_{1}, A_{2}$ are complex matrices, parameterized with $\vec{p}_1, \vec{p}_2$ variables correspondingly. The diagonal matrix $P$ includes phase shifts $\vec{\phi}$ or current-dependent phase functions $e^{i\phi_{i}(x)}$ (see Appendices \ref{app:ml-based-dataset-phases} and \ref{app:ml-based-dataset-currents} for details). Data sets $(\vec{\phi},U)$ and $(\vec{x},U)$ were divided into a training and a validation set. The training set was used to learn $A_{1,2}$ and $P(\vec{x)}$. The interferometer model was then evaluated using the validation set. The results are presented in Fig. \ref{fig:ML_histograms}a. As can be seen from the results, the fidelity between the measured unitary matrices and the predicted by both $(\vec{\phi},U)$ and $(\vec{x},U)$ models is no less than $0.90$ with a peak around $0.99$, which indicates the precision of the reconstructed digital interferometer model using the method of \cite{kuzmin2021architecture}.

However, the phase shifts-matrices dataset requires knowledge of the phase shift information, which is obtained from the calibration procedure that provides the mode mixing elements $M_{1,2}$ by itself. Therefore, performing another optimization procedure to determine the $A_{1,2}$ matrices, while already having the $M_{1,2}$ may seem somewhat redundant. In this regard, a more fundamental dataset of currents and corresponding unitary matrices for chips' transformation is in high demand.

It is worth mentioning that in this experiment, training the artificial network on a currents-matrices dataset ($\vec{x}$ - $U$) to determine the ML-based optical chip model aims to solve exactly the same task as the calibration of the phase shift layer. Both methods return the digital chip model as a function of the experimentally tunable parameters: electrical currents $\vec{x}$. This model enables the prediction of the unitary transformation of the chip at a given current. However, despite sharing the same goal, these two methods differ in the experimental data they utilize. The calibration method requires all the output power from the current dependencies, whereas the ML method requires a set of unitary matrices measured on the chip. Under realistic experimental conditions, it may not always be easy to perform full optical chip tomography, in contrast to simple output power calibration measurements. At least, because if we use the same experimental resources (coherent laser radiation), full chip tomography would require injecting the radiation into two ports of the chip simultaneously and precisely varying their relative phase shift. However, only one input port was used in the calibration measurements. Then both methods use optimization over a set of real-valued parameters to minimize the corresponding loss function.

\subsection{Optical port-to-port permutations in a wide range}
%The procedure described above for the experimental calibration of the photonic chip (and programming the chip at the same time) requires simply injecting coherent radiation into each mode of the chip, applying current to a specific heater, and recording the optical power from all output ports. 
After a successful approximation of the entire data set of the calibration data collected, a complete digital model of the chip is obtained. In this procedure, the wavelength $\lambda$ determines the resulting unitary transformation $U = U(\lambda)$ due to the spectral features of the mode-mixing blocks and the spectral dependence of the phase shifts. Reconstruction of the digital model can be repeated with different wavelengths $\lambda_{new}$ to retrieve the chip model $U=U(\lambda_{new})$. The multiport interferometer architecture \cite{Robust2020} was theoretically proven to be robust to the deviations of the mode-mixing block transfer matrices. By implementing the same operation, the optical port-to-port switching, using interferometers with different mode-mixing blocks, we highlight the robustness feature of the multiport architecture.

To demonstrate the robustness of our photonic chip to the imperfections in the mode-mixing layers and its potential to perform predefined optical transformations with practical applications, we have experimentally performed optical switching at seven different wavelengths between $915$ and $975$ nanometers. The optical switching is a particularly demanding task for an interferometric circuit, especially based on MZI blocks, because it requires precise calibration of the power couplers and phase shifters. We first calibrated our optical chip for seven wavelengths using a continuously tunable CW laser (Toptica CTL 950), which has a range of wavelengths between $\lambda_m=915$ and $975$. We then obtained digital models $U(\vec{x},\lambda_m)$ for each of these wavelengths. Then, using the known digital model of the chip, $U(\vec{x}, \lambda_m)$, we use optimization (the Python scipy.optimize.minimize function) to find a current configuration, $\{x_1, x_2, x_3\}$, which realizes a particular optical switch for each wavelength and each input port. Numerical optimization was used to minimize the infidelity function $Inf = 1 - F$ with fidelity defined as
\begin{equation}\label{eq::vector_fidelity}
    F(\vec{x}, \vec{y}) = \left( \sum_j \sqrt{x_j, y_j}  \right),
\end{equation}
where $\vec{x}$ and $\vec{y}$ are two column vectors, with a unit sum representing the normalized measured and target output power distributions at the output of the optical chip, respectively.

Finally, the optimized current configurations were set on the photonic chip, and the corresponding optical switching was observed. Figure\ref{fig:main_res_100_unittaries}c shows the average port-to-port switching fidelity among all switching configurations for each wavelength of coherent radiation. It can be seen that while the absolute values of the average port-to-port switching fidelities are not ideal (all less than 0.98) for each wavelength, the mutual average fidelity between the measured output power distribution and its prediction by a digital model is at least 0.99 (see the Appendix \ref{app:switch_details} for details). This clearly shows that if the target interferometer is unable to implement a required transfer matrix, the digital model provides an accurate estimation of the expected fidelity. 

\section{Discussion}

It should be noted that although the necessary currents for the realization of a particular unitary transformation $U(\vec{x}$ on the chip were obtained through optimization, this optimization process was not performed directly on the chip, as previously reported in \cite{kondratyev2024large}. Instead, it was conducted using a digital chip model. Therefore, using a digital model effectively saves resources of the target photonic chip and helps to extend its active lifetime by avoiding the need for excessive reconfiguration. Additionally, it can be shown that the scalability of the proposed calibration method increases quadratically with the number of required single heater calibrations on the optical chip (for details, see Appendix \ref{app:scaling}). This is a typical scaling factor for reconfigurable multi-port optical chip calibration routines.

Finally, we successfully demonstrate the implementation of $100$ random unitary matrices on a chip and the broadband optical port-to-port switching. These results validate our digital model and demonstrate that the optical chip with its complex interferometric structure was successfully programmed. 
In the future, we plan to demonstrate the implementation of Haar random unitaries on our chip and show the potential applicability of our chip in solving information processing tasks.

\begin{acknowledgements}
The authors acknowledge support from Russian Science Foundation grant 22-12-00353-P (https://rscf.ru/en/project/22-12-00353/); V.K., A.D. and G.G. acknowledge support of the Ministry of Science and Higher Education of the Russian Federation FSME-2025-0004 (PICs fabrication).   
\end{acknowledgements}

\bibliography{PLib}

\appendix
\section{Fabrication process \label{app:fab_process}}

\begin{figure*}[t]
\centering
\includegraphics[width=1\linewidth]{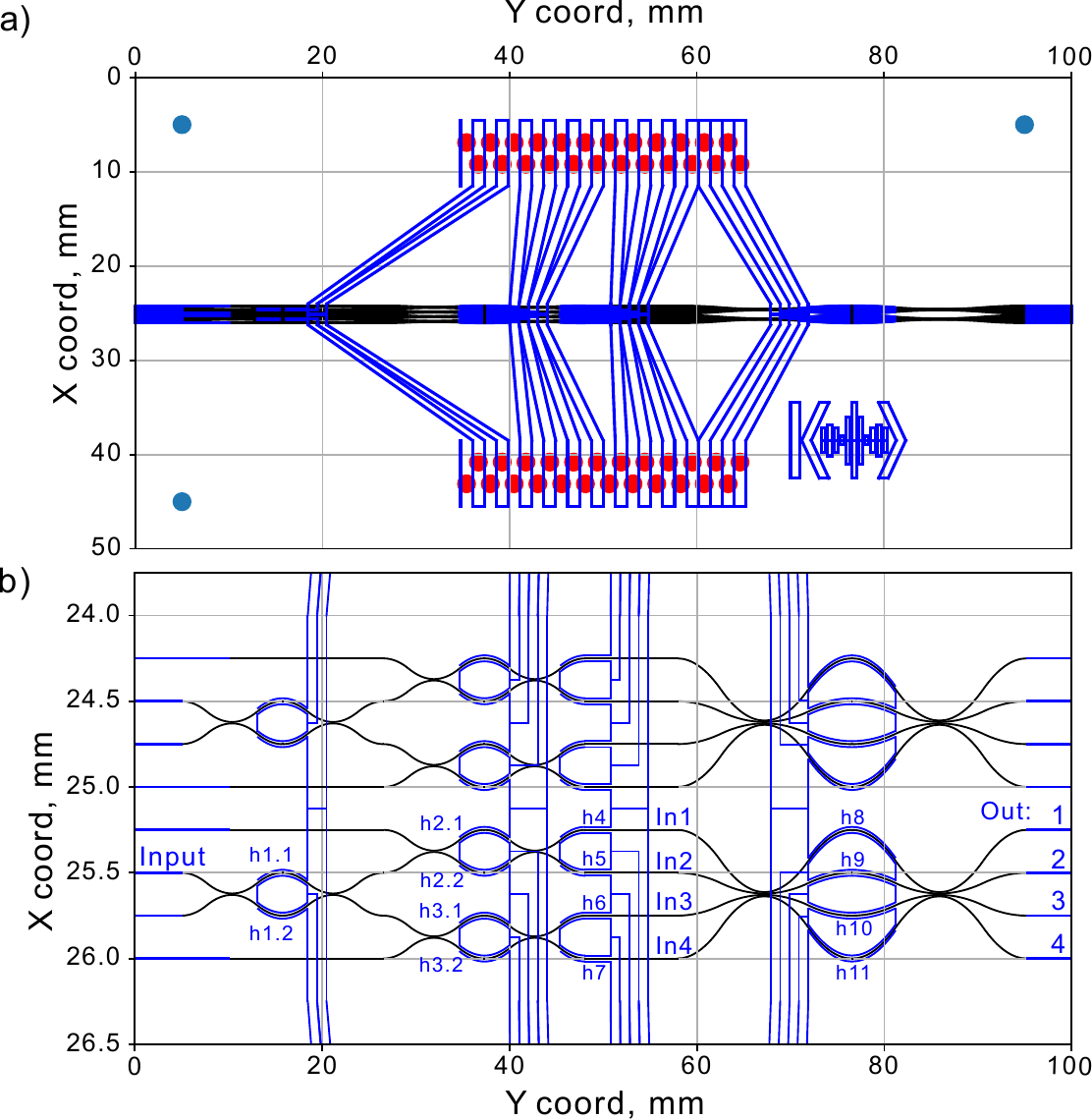}
\caption{ The actual optical chip structure, viewed from the top. a) A real-scale schematic of chip. The red and black dots represent electrical and ground connections to the PCB, respectively, and the blue dots represent special markers on a fused silica sample, which are required for precise chip alignment before electrode engraving. b) Zoomed-in part of waveguide structure and electrodes. Waveguides are depicted by black solid lines, and engraved electrodes by solid blue lines. A thin metal film covers the entire top surface of the chip, meaning that the entire white area in the figure is conductive, whereas the blue lines represent isolation trenches between electrodes. } 
\label{fig:electrodes}
\end{figure*}

The waveguide circuit was written using 515-nanometer laser pulses, which were the second harmonic of the Avesta Antaus ytterbium fiber femtosecond laser system. These pulses had a duration of 280 femtoseconds and were delivered at a repetition rate of 1 megahertz. Each pulse had an energy of 120 nanojoules and a linear polarization parallel to the writing direction inside the fused silica glass sample. The laser beam was focused using an aspheric lens with a numerical aperture (NA) of approximately 0.55, 15 microns below the surface of the sample. A 150-micrometer thick cover glass was placed between the lens and the sample to partially correct for spherical aberrations. A reconfigurable beam expander was used to enlarge the input aperture of the focus lens. A high-precision AeroTech Fiber Glide 3D air-bearing system was used to move the sample at a speed of 0.2 mm/s during the waveguide fabrication process.
The average propagation loss is approximately 0.65 dB/cm at 925 nm, and the coupling loss is approximately 1.5 dB per end face. Additionally, the bending loss was less than 0.1 dB/cm for a bending radius of 60 mm used in the experiment.

\section{Optical chip design \label{app:chip_design}}

Our photonic processor contained an auxiliary part of three cascaded Mach-Zhenders' interferometers (MZIs) and a target $4 \times 4$ interferometer consisting of three thermo optical phase shifter sandwiched between two $4 \times 4$ multiport directional couplers.
The design of the entire optical chip, including the waveguide structure and the electrode pattern, is shown in Fig. ~\ref{fig:electrodes}. The radius of curvature for almost all bends is 60 mm, which is a compromise between the bending loss effect and the overall size of the waveguide structure that can be achieved using our manufacturing process. The inner pair of waveguides in the $4 \times 4$ directional couplers has approximately a three times larger radius of curvature.
The input and output ports are spaced 250 micrometers apart and are connected to single-mode fiber optic arrays.

The thermo-optical phase shifters (a.k.a. electrodes or heaters) are fabricated using the same FSLW setup, with the same process used for writing waveguides. All the FSLW setup parameters for the thermo-optical phase shifters fabrication are identical, except for the sample translation speed, which is five times faster during the electrode engraving process (0.2 mm/s for waveguide writing and 1 mm/s for electrode engraving) to shorten the processing time. The FSLW setup for electrode engraving is less demanding, as we only need to remove metal from the border of the electrode. The laser power should be sufficient to engrave the electrode border. However, it should not be set too high as it can damage the surface of the optical chip. Both conditions are typically automatically met for the FSLW waveguide writing process, which is much more difficult to achieve and may require additional research \cite{abou2019femtosecond}. 
In our fabrication process the thermo-optical phase shifters are engraved on a thin metallic film, which is sputtered after the waveguide structure of the interferometer has been written. The thermo-optical phase shifters must be precisely placed exactly above the corresponding waveguides. We ensure  electrode position by using special markers (blue dots in Fig.~\ref{fig:electrodes}a) previously inscribed underneath the surface in the corners of the fused silica substrate during the waveguide writing process.
Thermo-optical phaseshifters are 35~$\mu$m wide and 3.7~mm long metal stripes, which repeat the corresponding interferometer arm. The transverse spacing (perpendicular to the waveguide axis) between the heaters was 250~$\mu$m. 
The NiCr 0.2~$\mu$m thick film covers the entire surface of the chip; therefore, in Fig. \ref{fig:electrodes} all of the white space is covered with metal, while the blue lines represent the engraved tracks, which isolate the electrodes from each other.
The electrodes are connected to a multi-channel digital computer controlled current source via a PCB interface with spring-loaded contacts, which contact points are schematically shown as red and black dots in Fig. \ref{fig:electrodes}a (each red dot corresponds to a single heater, whereas each black dot corresponds to a common round electrode).
The resistances of all MZI heaters are approximately 1000~$\Omega$, while the resistances of the target interferometers' heaters are 1700~$\Omega$, as they have greater length.
In the actual optical chip, two identical interferometric structures were fabricated in case one of them turned out to be unusable for some reason or if more than two heaters corresponding to a particular interferometric arm were burned out.

\section{Preparation part calibration \label{app:prep_MZI_calibration}}

\begin{figure}[h]
\centering
\includegraphics[width=1\linewidth]{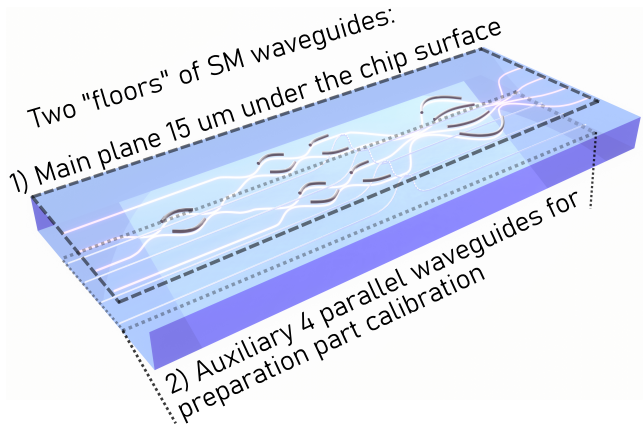}
\caption{ Schematic picture of the optical chips' actual waveguide structure. Four auxiliary parallel waveguides were used to calibrate the preparation part of the optical chip. Each of 4 auxiliary parallel waveguides weakly couples the corresponding output waveguide of the preparation part of the optical chip, which is sufficient for the calibration of each of three MZIs for realizing an arbitrary input to the target interferometer.  } 
\label{fig:chip_two_floors}
\end{figure}

The preparation part of our optical chip comprising three cascaded MZIs can be calibrated with the help of the four parallel auxiliary waveguides, which couples the output waveguides of the preparation part of the chip as shown in Fig. \ref{fig:chip_two_floors}. Therefore, to calibrate the three MZIs of the preparation part, we injected coherent laser radiation in the second input port of the chips' upper (and main) plane of the waveguides and registered the output radiation from the output ports of the auxiliary waveguides, i.e. a second (down) plane of the waveguide structure of our optical chip. 

In practice, coherent radiation is injected into the chip, and the output radiation is collected from the chip using v-groove fiber arrays placed on a six-axis micrometric positioner.

The optical power coupling between the preparation part of the interferometer and the auxiliary waveguides was deliberately weakened ($\leq 10\%$) to avoid losing much of the signal injected into the target interferometer. Meanwhile, the amount of coupled radiation was sufficient to calibrate the MZIs.

\section{Simultaneous approximation details \label{app:global_fit_details}}

As stated in the main text, the necessary data for calibration of the phase shift layer is obtained by consequently injecting coherent radiation into each input port of the target interferometer and sweeping current in some range at each heater while measuring the optical power from all the  output ports. 

Calibration data were gathered as follows: 

\begin{enumerate}
  \item Coherent laser radiation at wavelength $\lambda$ was injected in the $k$-th input mode of the target interferometer $k \in \overline{1, 4} $ (see Fig.~\ref{fig:experimental setup}b).
  
  \item The current applied to the heater $j \in \overline{1, 3}$ was applied in the range of $0$ to $15.6$~mA with a step of $0.12$ mA (resulting in $131$ different current values), while the optical power was measured from each $m$-th output port of the chip (see Fig.~\ref{fig:experimental setup}b). The measured optical power was then normalized. 

  \item The data was then organized in a way that allowed each data point to be accessed using four indices: input port number $k \in \overline{1, 4}$,  heater number $j \in \overline{1, 3}$ at which current $x$ was applied, index $i \in \overline{1, 131}$ that enumerates the applied current values $x_i$  ($x_0 = 0$ mA and $x_{131} = 15.6$ mA), and output port number $m \in \overline{1, 3}$ (there is no $m = 4$ as the output power was normalized $\sum_m P_{m} = 1$ and therefore there are three independent output powers left). 

  \begin{equation}\label{eq::data_point}
    P(k, j, i, m) = 
    \begin{Bmatrix}
     \textit{Normed power from $m$-th output port,} \\
     \textit{while laser is injected in $k$-th input port} \\
     \textit{and current $x_i$ is applied at $j$-th heater} \\
    \end{Bmatrix}
\end{equation}  
\end{enumerate}

Therefore, the complete experimental data contain $4716$ data points as $4 \text{input ports} \times 3 \text{heater} \times 3 \text{normed output ports}  \times 131 \text{current values}  = 4716$. Each data point $P_{exp}(k, j, i, m) \leq 1$ by construction.
 
\begin{figure*}[t]
\centering
\includegraphics[width=1\linewidth]{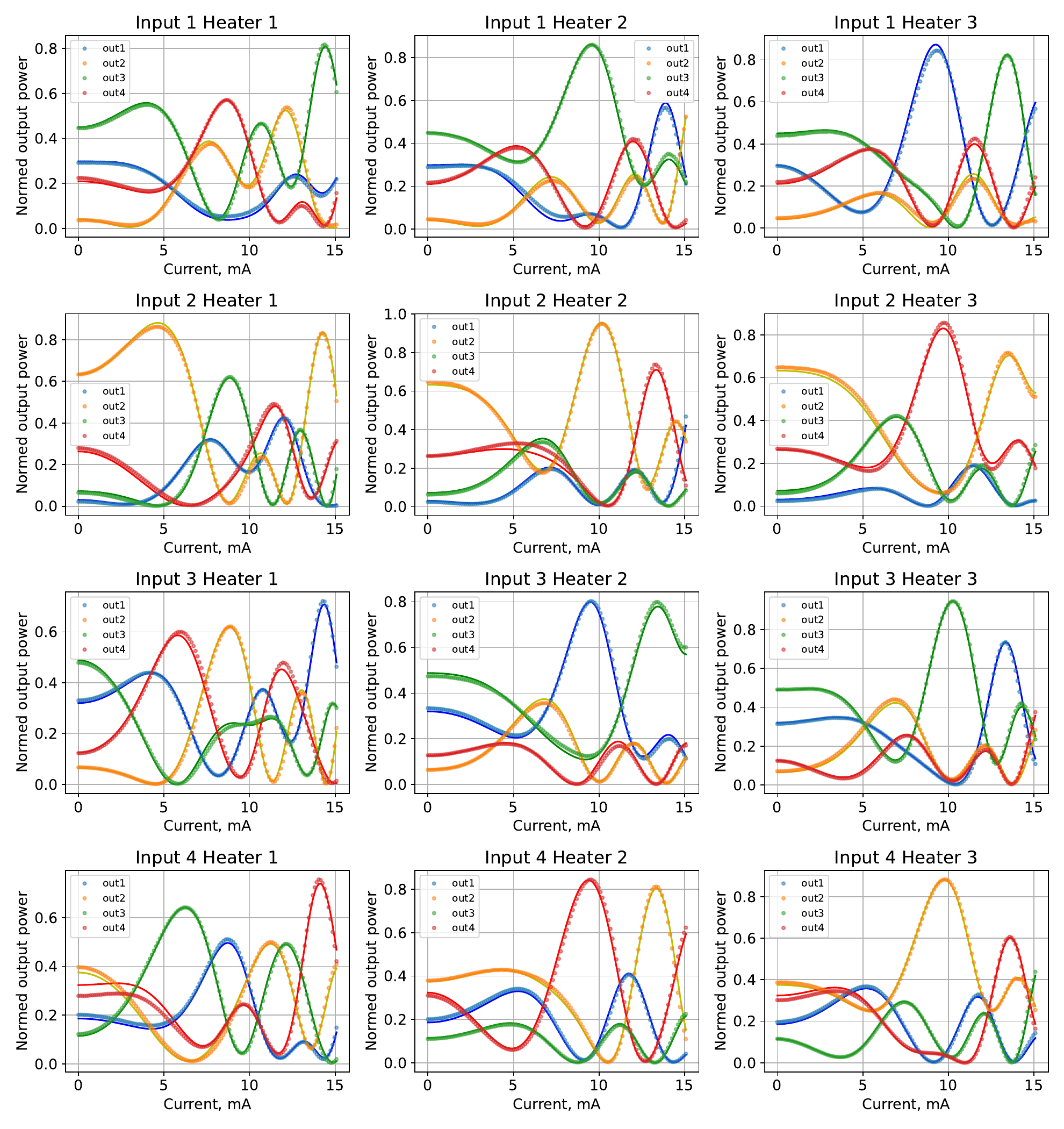}
\caption{ Simultaneous approximation of all the experimental calibration data according to (\ref{eq::power_with_crosstalk}). The measured normed optical power from the corresponding output port is depicted with dots, and a simultaneous approximation curve is shown as lines with corresponding colours. The coefficient of determination between the experimental data points and the simulated values $R^2 = 0.9972$. } 
\label{fig:big_fit_925}
\end{figure*}

An example of the experimentally measured complete calibration data for the wavelength $\lambda = 925$ nm is depicted using dots in Fig. \ref{fig:big_fit_925}. 

The digital model of the chip $U_{sim}$ can predict the normalized output optical power of the target interferometer corresponding to a concrete set of indices $\{k, j, i, m \}$ indices: $P_{sim}(k, j, i, m)$.
The main task of calibrating the phase shifters is to find a digital model of the target interferometer that approximates all experimentally measured data points $P_{sim}(k, j, i, m) \approx P_{exp}(k, j, i, m)$. This task is accomplished by optimizing the parameters of the chips' model $U_{sim}$ to maximize the $R^2$ coefficient of determination (loss function) as follows:
\begin{equation}\label{eq::curve_fit_R_squared}
    R^2 = 1 - \dfrac{ \sum_{k,j,i,m} ( P_{exp}(k, j, i, m) - P_{sim}(k, j, i, m) )^2 }{ \sum_{k,j,i,m} ( P_{exp}(k, j, i, m) - \overline{P_{exp}} )^2 },
\end{equation}
where $ \overline{P_{exp}} = \dfrac{1}{N} \sum_{k,j,i,m} P_{exp}(k, j, i, m)$ and $N$ is the total number of data points, which is equal to $N = 4716$ in our experiment. Thus, the described optimization procedure is generally a curve fitting. Further, we need to determine $30$ real parameters: $9$ in each mode mixing matrix $M_{1,2}$, $3$ bias zero phase shifts $\Phi_0$ and $9$ elements of the $3\times 3$ crosstalk  matrix $A = \{ \alpha_{ij} \}$. As the mode mixing matrices $M_{1,2}$ were parameterized by a universal interferometer consisting of triangular MZI meshes, all optimization parameters except for nine $\{ \alpha_{ij} \}$ are phase shifts that can be varied in the $[0, 2\pi]$ range. We have not supposed any a priori information about the form of $M_{1,2}$ and treat them as two independent generally different unitary matrices parameterized by nine real parameters  each ($t_{1-9}$ and $t_{10-18}$), whose initial values for optimization were chosen at random from $[0, 2\pi]$. The initial values for the bias phase shifts, $Phi_0$, are always taken as zero. However, the initial guess for the nine crosstalk matrix elements $\{ \alpha_{ij} \}$ should have been chosen more accurately, both because it did not have finite bounds of its' values as phase shift parameters $t_j$ and $\Phi_0$ had a $[0, 2\pi]$, and because it had an explicit physical meaning of thermal distribution in the sample material that must be governed by the geometry of the heaters and the heat equation. Therefore, a preliminary estimate of (at least) the relative values of the crosstalk matrix elements $\{ \alpha_{ij} \}$ is necessary.

\begin{figure*}[t]
\centering
\includegraphics[width=1\linewidth]{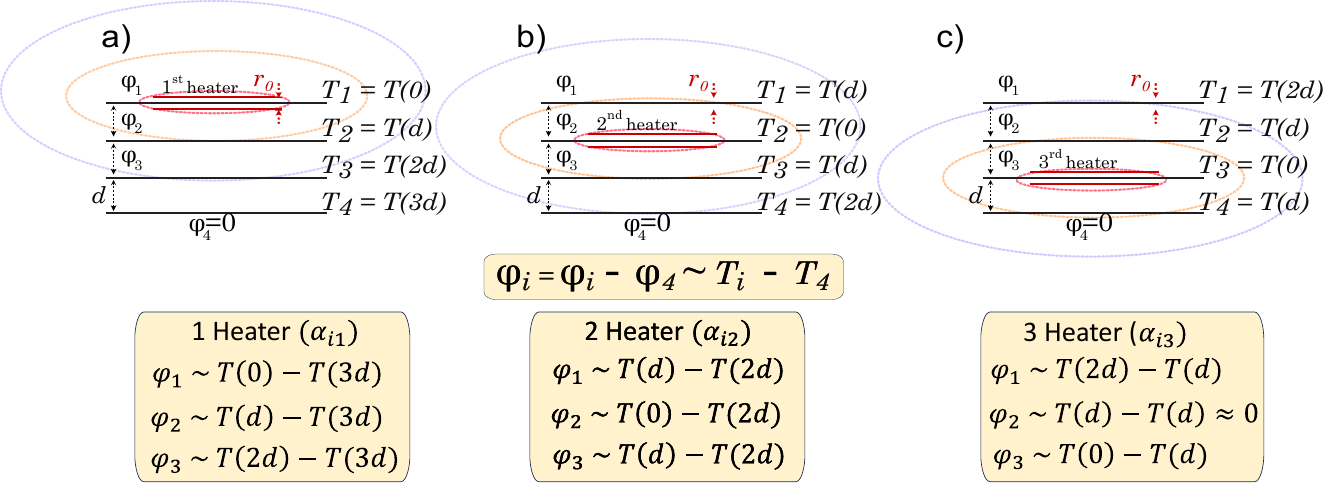}
\caption{  Guidelines for the relative crosstalk estimation. } 
\label{fig:crosstalk_appendix}
\end{figure*}

To estimate the relations between crosstalk matrix elements $\alpha_{ij}$, we considered a simple system of four equivalent (endless) straight and parallel heaters of width $r_0$ that are placed above four equivalent (endless) straight and parallel waveguides spaced at a distance  $d$ apart from each other (see Fig. \ref{fig:crosstalk_appendix}). Such a model system, perhaps, roughly describes the heaters in our actual target interferometer which are notably not straight, as shown in Fig. \ref{fig:electrodes}b. Nevertheless, this simplified model can still be helpful for estimation.

The physical meaning of the crosstalk matrix $A = \{ a_{ij} \}$ is how "strong" the $j$-th heater affects the $i$-th heater (or, more precisely, the $i$-th interferometric arm). For example, if current $x_1$ is applied to the first heater (see Fig. \ref{fig:crosstalk_appendix}a) it will cause the phase shift change not only in the first waveguide $\varphi_1(x_1)$, which it is supposed to affect, but in the second $\varphi_2(x_1)$ and third $\varphi_3(x_1)$ waveguides as well. However, the impact of the first heater on the second waveguide is less than that of the first one, and the impact of the first heater on the third waveguide is even lower than that of the second. This is because the phase shift induced by the heater is proportional to the temperature of the material in which the waveguide is embedded, and the temperature decreases with the distance from the thermal energy source. 

The known geometry of the system can be used to quantitatively estimate  the relations between $a_{ij}$ values.
Let the $j$-th heater receive some current $x$. 
% The working heater would change the temperature distribution in nearby areas that would cause the change in phase shifts in the adjacent waveguides.
We consider the logarithmic temperature decay with distance from the  heater:
\begin{equation}\label{eq::temperature_log}
    T(r)= 
\begin{cases}
     T_0, &  |r| \leq r_0\\
     T_0 - \beta \ln \dfrac{r}{r_0}, &  |r| > r_0,
\end{cases}
\end{equation}
where $T_0$ is the maximum temperature of the heater and $\beta$ is a constant with the dimension of the temperature.
For the four waveguides in the phase shift layer, three relative phase shifts are sufficient. We assume that all three phase shifts are relative to the fourth waveguide: $\varphi_i \equiv \varphi_i - \varphi_4$ for $i \in [1, 3]$, which means that $\varphi_i$ is the phase shift between the $i$-th waveguide and the fourth waveguide (or $\varphi_4 \equiv 0$).
The phase shift $\varphi_i$ arises due to the temperature difference between the $i$-th and fourth waveguides: $\varphi_i - \varphi_4 \sim T_i - T_4$. Therefore, when the $j$-th heater receives some current $x$, it leads to a phase shift $\varphi_{ij}$ in the $i$-th waveguide due to the temperature distribution in the material, as follows:
\begin{equation}\label{eq::crosstalk_temperature_distribution}
\begin{matrix}
    \text{1st heater} (\alpha_{i1}) & \text{2nd heater} (\alpha_{i2}) & \text{3rd heater} (\alpha_{i3}) \\
    \varphi_{11} \sim T(0) - T(3d) & \varphi_{12} \sim T(d) - T(2d) & \varphi_{13} \sim T(2d) - T(d) \\
    \varphi_{21} \sim T(d) - T(3d) & \varphi_{22} \sim T(0) - T(2d) & \varphi_{23} \sim T(d) - T(d) \\
    \varphi_{31} \sim T(2d) - T(3d) & \varphi_{32} \sim T(d) - T(2d) & \varphi_{33} \sim T(0) - T(d)
\end{matrix},
\end{equation}
, as shown schematically in Fig. \ref{fig:crosstalk_appendix}. Without concretizing the explicit dependence of the $T(r)$ it immediately follows from (\ref{eq::crosstalk_temperature_distribution}) that in the system with the simple geometry of parallel heaters, $\varphi_{23} \approx 0$, as the temperatures at the fourth and second waveguides are equal when current is applied to the third heater (which is due to the symmetry of the simplified system). Moreover, when the third heater was turned on, the temperature in the fourth waveguide was higher than that in the first, which led to the negativeness of $\varphi_{13} < 0$. Other qualitative relations between the induced phase shift $\varphi_{ij}$ can be excluded from (\ref{eq::crosstalk_temperature_distribution}) such as: $\varphi_{11}>\varphi_{22}>\varphi_{33}>\varphi_{21}>\varphi_{12} \approx \varphi_{32}>\varphi_{31}>\varphi_{23} \approx 0 > \varphi_{13}$, and $|\varphi_{13}| \approx \varphi_{12}$, which all follow from the fact that the temperature decreases with distance from the energy source.

The relations between the phase shifts $\varphi_{ij}$ become quantitative after substituting the logarithmic law for the temperature in (\ref{eq::crosstalk_temperature_distribution}):
\begin{equation}\label{eq::crosstalk_mat_estim_sub}
\begin{matrix}
    \text{1st heater} (\alpha_{i1}) & \text{2nd heater} (\alpha_{i2}) & \text{3rd heater} (\alpha_{i3}) \\
    \varphi_{11} \sim \beta \ln\dfrac{3d}{r_0} & \varphi_{12} \sim \beta \ln 2 & \varphi_{13} \sim -\beta \ln 2 \\
    \varphi_{21} \sim \beta \ln 3 & \varphi_{22} \sim \beta \ln\dfrac{2d}{r_0} &  \varphi_{23} \approx 0 \\
    \varphi_{31} \sim \beta \ln \dfrac{3}{2} & \varphi_{32} \sim \beta \ln 2 & \varphi_{33} \sim \beta \ln\dfrac{d}{r_0}
\end{matrix},
\end{equation}
from which, after dividing everything by the constant $\ln{\dfrac{3}{2}}$ the crosstalk matrix $A$ can be written as:
\begin{equation}\label{eq::crosstalk_relation_appendix}
    \centering
    A_{model} = \alpha_0
    \begin{pmatrix}
    \log_{1.5} { \dfrac{3d}{r_0}} & 1.71 & -1.71 \\
    2.71 & \log_{1.5}{ \dfrac{2d}{r_0}} & 0 \\
    1 & 1.71 & \log_{1.5}{ \dfrac{d}{r_0} } \\
    \end{pmatrix},
\end{equation}
where $1.71 = \log_{1.5} 2$ and $2.71 = \log_{1.5} 3$.
As can be seen from (\ref{eq::crosstalk_relation_appendix}) the relations between the non diagonal elements of $A$ are constant for any system of straight parallel heaters (and waveguides), whereas the diagonal elements of $A$ are functions of a dimensionless parameter $\xi = \dfrac{d}{r_0}$.

To estimate the initial guess for the real crosstalk matrix $A_exp$ describing the actual geometry of the heaters (see Fig. \ref{fig:electrodes}b) we substitute in (\ref{eq::crosstalk_relation_appendix}) the approximate value of $\xi$ relevant for our device. In our target interferometer, the geometries of all heaters are arcs of a circle. The effective parameter $d_{exp}$ of our actual chip can be taken as the average between the largest $d_{max} = 250$ $\mu$m in the center of a phase shift layer and the lowest $d_{min} \approx 130$ $\mu$m at the edges of the heaters, which gives $d_{exp} \approx 190$ $\mu$m. The width of the heater was the same for all heaters and was equal to $r_0 = 35$ $\mu$m. Therefore, for $\xi_{exp}$ we have $\xi_{exp} = \frac{d_{exp}}{r_0} \approx 5.43$, resulting in an estimation of the initial guess for the crosstalk matrix $A_{estim}$:
 \begin{equation}\label{eq::crosstalk_relation_estmation}
    \centering
    A_{estim} = \alpha_0
    \begin{pmatrix}
    6.88 & 1.71 & -1.71 \\
    2.71 & 5.88 & 0     \\
    1    & 1.71 & 4.17  \\
    \end{pmatrix},
\end{equation}

With (\ref{eq::crosstalk_relation_estmation}) as an initial guess for $A$, three zeros as a starting point for $\Phi_0$ and $18$ random numbers from $[0, 2\pi]$ for $M_{1,2}$, we launched optimization and successfully converged to a value of the coefficient of determinacy $R^2 = 0.9972$ at $4716$ data points. The predicted output powers of the model $P_{sim}(k, j, i, m)$ and the experimentally measured data points $ P_{exp}(k, j, i, m)$ are plotted in Fig. \ref{fig:big_fit_925} as curves and dots, respectively.

The successfully terminated optimizer returned the following values for $A_{exp}$:
\begin{equation}\label{eq::crosstalk_relation_experimental}
    \centering
    A_{exp} = \alpha_0^{exp}
    \begin{pmatrix}
    6.79 & 1.5 & -2.0 \\
    2.71 & 6.0 & -0.29     \\
    1    & 1.71 & 4.29  \\
    \end{pmatrix}.
\end{equation}
As we can see from the optimization crosstalk matrix $A_{exp}$ did not go too far from the initial estimation $A_{estim}$, which demonstrates the goodness of the simplified model used for the crosstalk matrix estimation $A_{estim}$. Nevertheless, a qualitative difference between $A_{exp}$ and $A_{estim}$ can be observed that is slightly negative $\alpha_{23} < 0$. However, this value can be legitimately explained by the actual geometry of the heaters and waveguides of our chip (see Fig. \ref{fig:electrodes}b). In fact, the distance between the third heater and the fourth waveguide is always slightly smaller (or equal at the center) than the distance between the third heater and the second waveguide, resulting in higher temperatures at the fourth waveguide compared to the second, leading to a negative phase shift in the second waveguide relative to the fourth.

No special assumptions on phase shift bias  (zero current) $\Phi_0$ were made and their initial guess values were set to zero $\Phi^{init}_0$. The Optimized values for $\Phi^{exp}_0$ at a wavelength 925 nm appeared to be $\Phi^{exp}_0 = \{ 0.23, 0.98, 1.02 \}$ rad. These values can also be explained by the actual geometry of the waveguides in the phase shift layer, as the first and fourth waveguides have the same length $L_{1,4}$ and the second and third waveguides have the same length, which is clearly smaller than $L_{1,4}$ (see Fig. \ref{fig:electrodes}b).

\section{Unitary transformation of the optical chip measurement  \label{app:unitaries_measure}}

The preparation interferometer, which consisted of three cascaded MZIs, was mainly used as a chip-based instrument for accurate and automated unitary transformation measurement of the target interferometer $U$.
Measurement of the unitary matrix of the target interferometers was performed by exploiting coherent laser radiation according to \cite{rahimi2013direct}.

The task of the experimental measurement of the unitary matrix $U_{ij} = |U_{ij}|e^{i\phi_{ij}} $ is to determine all its' modules $|U_{ij}|$ and angles $ \phi_{ij}$.
Modules $|U_{ij}|$ of the target unitary can be obtained simply by injecting  coherent laser radiation into each single input port of the target interferometer and then measuring the normalized power distribution $P_{ij}$ from all output ports of the target interferometer:
\begin{equation}\label{eq::unitary_modules}
    |U_{ij}| = \sqrt{ \frac{P_{ij}}{\sum_j P_{ij}} },
\end{equation}
where $P_{ij}$ is the measured optical power from the $i$-th output port of the interferometer, whereas radiation is injected only into the $j$-th input port. Repetition of this experimentally simple procedure $N$ times for each input port of the target interferometer, a matrix of its modules $|U_{ij}|$ can be obtained.
However, despite the simplicity of this measurement, there is little subtlety in determining the modules $|U_{ij}|$ of a linear optical interferometer. This arises from the possible presence of inhomogeneous losses at the output ports of the target interferometer, which would lead to the non uniform sum of the absolute values squared of the matrix elements $S^{row}_i = \sum_j|U_{ij}|^2 \neq 1$ that should sum to $1$ for general unitary matrices. Whereas, all columns sum to $1$ by construction $S^{col}_j = \sum_i|U_{ij}|^2 = 1$. This inconsistency can be resolved by adjusting the matrix of squared modules $|U_{ij}|^2$ multiplying it from both sides by two diagonal matrices $D_1$ and $D_2$ with strictly positive elements, representing losses at input and output of the chip, such that the new matrix $|U_{ij}|^2_{new}$ satisfies the unitary requirements and all its' rows and all columns sum to $1$. These matrices $D_1$ and $D_2$ can be found by a straightforward iterative method called the Sinkhorn–Knopp algorithm \cite{sinkhorn1967concerning}. Therefore, this rescaling of the modules initially measured $|U_{ij}|$ should be performed if necessary.

Angles $\phi_{ij}$ can also be extracted from optical power measurement by registering the interference that requires injecting coherent laser radiation into two input ports. 

We assume that coherent radiation is injected into the $l_1$ and $l_2$ input ports of the target interferometer in the form:
\begin{equation}\label{eq::angles_measure_input_state}
    E_{in} =  \{ \underset{1}{0}, \dots,  \underset{l_1}{a}, \dots, \underset{l_2}{b e^{i\theta}}, \dots, \underset{N}{0}\}^T,
\end{equation}
where $a$ and $b e^{i\theta}$ are the complex amplitudes of the input coherent field such that $a^2 + b^2 = 1$. Without loss of generality, assume that $a$ and $b$ are real. Therefore, $\theta$ is their relative phase shift. Let us also introduce a parameter $x$ that can control the value of $\theta$. This parameter plays a crucial role in the measurement of angles $\phi_{ij}$, as in a typical experiment, coherent radiation split into two beams, which are then injected into the $l_1$ and $l_2$ input ports of the target interferometer. Then, to one of the split beams an extra phase shift $\theta(x)$ is added by varying the experimental parameter $x$ and the output optical power distribution $P(x)$ among the output ports of the target interferometer is measured. The angles of the $l_2$ column of the target matrix $\phi_{il_2}$ can be extracted from the power measured from the $x$ curves $P_k^{l_1,l_2}(x)$. To determine angles in all other columns, usually $l_2$ is varied from $1$ to $N$ except for $l_2 \neq l_1$, while $l_1$ is fixed.

It can be shown that while radiation is injected into the $l_1$ and $l_2$  input ports, the optical power from the $k$-th output port of the target interferometer as a function of $x$ is governed by:
\begin{equation}\label{eq::angles_measure_input_P_law}
    P_k^{l_1,l_2}(x) = C_{l_1, l_2} + D_{l_1, l_2}  \cos(\theta(x) + \phi_{kl_2}),
\end{equation}
where $C_{l_1, l_2}$ and $D_{l_1, l_2}$ are some constants (depending on the modulus of $U$). It should be noted that in (\ref{eq::angles_measure_input_P_law}) all angles in the $l_1$-th column of the $U$ matrix were assumed to be zero ($\phi_{k l_1} \equiv 0$ for each $k \in \overline{1,N}$), as they can be associated with $N$ irrelevant phase shifts $P_{out}(\phi_{k l_1})$ after the target interferometer $ \Tilde{U} \equiv P_{out}(\phi_{k l_1}) \times U$ \cite{rahimi2013direct}.

Then, after obtaining the experimental curves $P(x)$ they can be approximated according to (\ref{eq::angles_measure_input_P_law}) and the necessary angles $\phi_{kl_2}$ can be determined for each $k \in \overline{1,N}$ and each $kl_2 \in \overline{1, N-1}$ ($l_2$ varies from $1$ to $N-1$ because $l_2$ is a number corresponding to one of the two irradiated input ports of the target interferometer, and the other is $l_1$). 

We realized the procedure described above on a chip with the help of a preparation interferometer that can prepare any desired coherent input state to be sent to the target interferometer, including those of the form (\ref{eq::angles_measure_input_state}). For the measurement of angles of the unitary $U$, we used thermo-optical heaters at the output of the preparation part. Therefore, in our case, the current served as the $x$  parameter and $\theta(x) = \theta_0 + \alpha_{\theta} x^2 $ leading to the $P_k^{l_1,l_2}(x)$  dependence of the form:
\begin{equation}\label{eq::angles_measure_input_P_law_OUR}
    P_k^{l_1,l_2}(x) = C_{l_1, l_2} + D_{l_1, l_2}  \cos(  \alpha^{l_2} x^2 + \gamma_{kl_2} ),
\end{equation}
where$ \gamma_{kl_2} = (\theta^{l_1, l_2}_0 + \phi_{kl_2})$.

Therefore, to determine the angles of the unitary transformation of the target interferometer, we measured all the necessary $P_k^{l_1,l_2}(x)$ curves and approximated each using ( \ref{eq::angles_measure_input_P_law_OUR} ), which returns the values of the corresponding shifts in the sine angles $\gamma_{kl_2}$. For fixed $l_1$ and $l_2$ each angle $\gamma_{kl_2}$ has the extra term $\theta^{l_1, l_2}_0$, which can be eliminated by subtracting angle $\phi_{l_1l_2}$ from each $\gamma_{kl_2}$ having
$(\Tilde{ \gamma}_{k l_2} = \phi_{k l_2} - \phi_{l_1 l_2})$ for each $k \in \overline{1, 4}$ and each $l_2 \in \overline{1, 4}, l_2 \neq l_1$.

Eventually, the procedure for the target unitary reconstruction $U$ results in the experimentally determined matrices of modules $|U_{ij}|$ and angles $\Tilde{\gamma}_{ij}$.
The unitary matrix $\Tilde{U}_{ij}$ of interest is constructed using element-wise matrix multiplication as follows:
\begin{equation}\label{eq::angles_measure_final_U}
    \Tilde{U}_{ij} = |U_{ij}| \cdot e^{ i \Tilde{\gamma}_{ij} },
\end{equation}
where $2N - 1 = 7$ values of $\Tilde{\gamma}_{ij}$ have zero values by construction. These are the angles of the $l_1$-th column and the $l_1$-th row of  the reconstructed matrix $\Tilde{U}$ on the chip. The fact that in the reconstructed unitary transformation $\Tilde{U}$ of the target $N$-port interferometer,  $2N - 1$ angles are set to be real is inherent to any tomography algorithm  \cite{laing2012super, rahimi2013direct, heilmann2015novel, suess2020rapid}. This is due to the physical equivalence of $U$ and  $\Tilde{U} = P_{out}(\vec{\phi_{out}}) \times U \times P_{in}(\vec{\phi_{in}})$, where $P_{in}(\vec{\phi_{in}})$ and $P_{out}(\vec{\phi_{out}})$ are diagonal matrices with pure phases $\vec{\phi_{in}}$ and $\vec{\phi_{out}}$, respectively. 
In other words, any tomography algorithm can reconstruct the unitary transformation of a linear optical interferometer only up to the input and output phases \cite{laing2012super, rahimi2013direct, heilmann2015novel, suess2020rapid}.

As a final remark on the reconstruction procedure, the modules and the angles measurements are conducted completely independently, which does not provide any guaranty that the reconstructed matrix by (\ref{eq::angles_measure_final_U}) $\Tilde{U}$ is unitary. Even in the case where one had performed the Sinkhorn-Knopp algorithm \cite{sinkhorn1967concerning} to rescale the  experimentally measured modules of $\Tilde{U}$, the angles were still obtained using a separate interferometric procedure. However, if $\Tilde{U}$ appeared to be nonunitary, it can be projected onto the closest matrix in the unitary subspace by, for example, polar decomposition or single value decomposition \cite{hall2013lie}.

As, typically, optical power from all output ports of the chip is collected simultaneously each port by each photo detector, the number of separate measurements necessary to obtain the modules $|U_{ij}|$ for a $N\times N$-port interferometer is equal to $N$. However, angle measurement $\Tilde{\gamma}_{ij}$ requires the collection of $N-1$ interference fringes, resulting in a total number of measurements for the unitary matrix $N\times N$ reconstruction of $2N - 1$.

In our experiment, we set $l_1 = 1$ for every measurement of  $\Tilde{U}$ on the chip, which makes the first row and the first column of any measured $\Tilde{U}$ real.
However, because the measured unitary matrix of the target interferometer $\Tilde{U}$ is in the form of a real first row and first column, the simulated matrix $U_{sim}$, generated by the chips' digital model, should be converted to this same form before their comparison, since the matrix fidelity (\ref{eq::matrix_fidelity_form}) measure is sensitive to such deviations, while these unitary matrices are physically equivalent.
This type of unitary matrix $U_{sim} = \rho_{ij} \exp(i \phi_{ij})$ transformation can be simply performed by multiplying it on both sides by diagonal unitary matrices:
\begin{equation}\label{eq::real_border_1}
    \Tilde{U}_{sim} = \Gamma \times  U_{sim} \times \Omega,
\end{equation}
where $\Gamma = diag( e^{i \gamma_j} )$ and $\Omega = diag( e^{i \omega_i} )$, the elements of which are equal to $\gamma_j = -\phi_{1j} + \phi_{11}$ and $\omega_i = - \phi_{i1}$. After multiplying by $\Gamma$ and $\Omega$, the unitary matrix $U_{sim}$  transforms to $ \Tilde{U}_{sim}  = \rho_{ij} \exp(i \Tilde{\phi}_{ij})$ with zero angles in the first row and first column by construction: $\Tilde{\phi}_{1j} \equiv \Tilde{\phi}_{i1} \equiv 0$.

More generally, any $N \times N$ unitary matrix $U_{ij} = \rho_{ij} \exp(i \phi_{ij})$ can always be multiplied on both sides by diagonal $N \times N$ unitary matrices $\Gamma = \text{diag}( e^{i \gamma_j} )$ and $\Omega = \text{diag}( e^{i \omega_i} )$ with appropriate angles $\gamma_j$ and $\omega_i$ such that the resulting unitary matrix $\Tilde{U}_{ij} = \rho_{ij} \exp(i \Tilde{\phi}_{ij})$ has real $I$-th row and $J$-th column: $\Tilde{\phi}_{iJ} \equiv \Tilde{\phi}_{Ij} \equiv 0$. The necessary angles $\gamma_j$ and $\omega_i$ are obtained from:
\begin{equation}
    \omega_i = -\phi_{iJ}, \ \
    \gamma_j = -\phi_{Ij} + \phi_{IJ}.
\end{equation}\label{eq::real_border_2}

We call this operation \emph{ the real border} the unitary matrix and define a \emph{real border} function $f_{rb}(U, I, J)$ that takes a unitary matrix $U$ at its' input and returns another unitary matrix $\Tilde{U}$ with a purely real $I$-th row and $J$-th column.

Therefore, before calculating the matrix fidelity function (\ref{eq::matrix_fidelity_form}) between the experimentally measured unitary transformation of the chip $U_{exp}$ and the corresponding matrix $U_{sim}$ simulated by the chips' digital model, both should be equivalently \emph{real bordered}. All matrix fidelities shown in the histogram in Fig. \ref{fig:main_res_100_unittaries}a were calculated between appropriately \emph{real bordered} pairs of $U_{exp}$ and $U_{sim}$.

\section{Phase shifts-matrices ($\vec{\varphi}$-$U$) dataset}\label{app:ml-based-dataset-phases}

The data set contained $27$ pairs of unitary matrices and the corresponding phase shifts $\vec{\varphi}$ obtained from the experimental calibration. This means that, in this data set, we have the knowledge of the exact phase shifts $\vec{\varphi}$ that were set on the chip and the corresponding measured unitary transformation $U_{data}(\vec{\varphi})$ of the chip.
The neural network was trained on $27$ unitary matrices for which the fidelity between them and the calibration model is greater than $0.95$ (see Fig. \ref{fig:main_res_100_unittaries}a) and tested on the other $65$ measured unitary matrices.

The trained model was:
\begin{equation}\label{eq::training_model_1}
    U_{trained}(\vec{p}) = f_{rb}(  A_2(\vec{p}_2) \times P(\vec{\varphi}) \times A_1(\vec{p}_1)  ),
\end{equation}
where $ f_{rb}(\cdot)$ is a \emph{real border} function defined in Appendix \ref{app:global_fit_details}, $A_{1,2}$ is two arbitrary complex matrices (not necessarily unitary), and $ P(\vec{\varphi}) = diag(e^{i \varphi_1}, e^{i \varphi_2}, e^{i \varphi_3}, 1 )$ is a diagonal phase shift matrix. 
The matrices $A_{1,2}$ were considered as general complex valued $4\times 4$ matrices and were parameterized by $2\times 4^2 = 32$ real parameters $\vec{p}$ each, representing the parameterization of the real and imaginary parts of $A_{1,2}$. The initial values of $\vec{p}$ were chosen at random for each optimization run.

We used the Adaptive Moment Estimation optimization algorithm (Adam) with $5000$ epochs with mini batches containing $5$ random elements each. The Frobenius metric was chosen as the loss function, as in \cite{kuzmin2021architecture}.

The results are presented in Fig. \ref{fig:ML_histograms}a. As can be seen from the results, the fidelity between the measured unitary matrices and predicted by the ML model is no less than $0.90$ with a peak around $0.99$, which indicates the accuracy of the revealed ML-based optical chip model.
However, the phase shifts - matrices data set requires exact knowledge of the phase shifts information, which is obtained from the calibration procedure that provides the mode mixing elements $M_{1,2}$ itself. Therefore, performing another optimization procedure to determine the $A_{1,2}$ matrices, while already having the $M_{1,2}$ may seem somewhat redundant. In this regard, a more fundamental dataset of currents and corresponding unitary matrices for chips' transformation is in high demand.

\section{Currents-matrices ($\vec{x}$ - $U$) dataset}\label{app:ml-based-dataset-currents}

The data set contained $60$ pairs of unitary matrices and the corresponding currents $\vec{x}$ set on the chip. This scenario is of more practical importance than the $\vec{\varphi}$ - $U$ data set, since it contains only raw primal data without any previous processing, such as the calibration routine.

The neural network was trained on $60$ random unitary matrices from measured matrices on the optical chip (see Fig. \ref{fig:main_res_100_unittaries}a) and tested on the other $36$ measured unitary matrices. In contrast with data set $\vec{\varphi}$ - $U$ in data set $\vec{x}$ - $U$, we assume that there is no calibration information $\varphi(x)$ and therefore cannot choose $60$ unitary matrices with calibration model fidelity greater than $0.95$ as in the previous subsection.

The trained model was:
\begin{equation}\label{eq::training_model_2}
    U_{trained}(\vec{p}) = f_{rb}(  A_2(\vec{p}_2) \times P(\vec{x}) \times A_1(\vec{p}_1)  ),
\end{equation}
where $ f_{rb}(\cdot)$ is a real border function, $A_{1,2}$ is two arbitrary complex matrices (not necessarily unitary), and $ P(\vec{x}) = diag(e^{i \varphi_1(\vec{x})}, e^{i \varphi_2(\vec{x})}, e^{i \varphi_3(\vec{x})}, 1 )$ is a diagonal phase shift matrix with phase shifts from current dependence (\ref{eq::phi_with_crosstalk_matrix}), resulting in $12$ real parameters: $3$ bias phase shifts $\vec{\varphi}_0$ and $9$ crosstalk matrix elements $A_{ij}$. The matrices $A_{1,2}$ were again considered to be general complex-valued  matrices $4\times 4$ and were parameterized by $32$ real parameters. The initial values for $\vec{p}$ were chosen at random for each optimization run, while an initial guess for the crosstalk matrix elements $\{ \alpha_{ij} \}$ was taken according to the estimations discussed in the previous section (see Appendix \ref{app:global_fit_details}).

To train the network, the Adaptive Moment Estimation optimization algorithm was used with $6000$ epochs with mini batches containing $5$ random elements each. The Frobenius metric was chosen as the loss function as in \cite{kuzmin2021architecture}.

\section{Broadband optical switching details}\label{app:switch_details}

\begin{figure*}[t]
\centering
\includegraphics[width=0.9\linewidth]{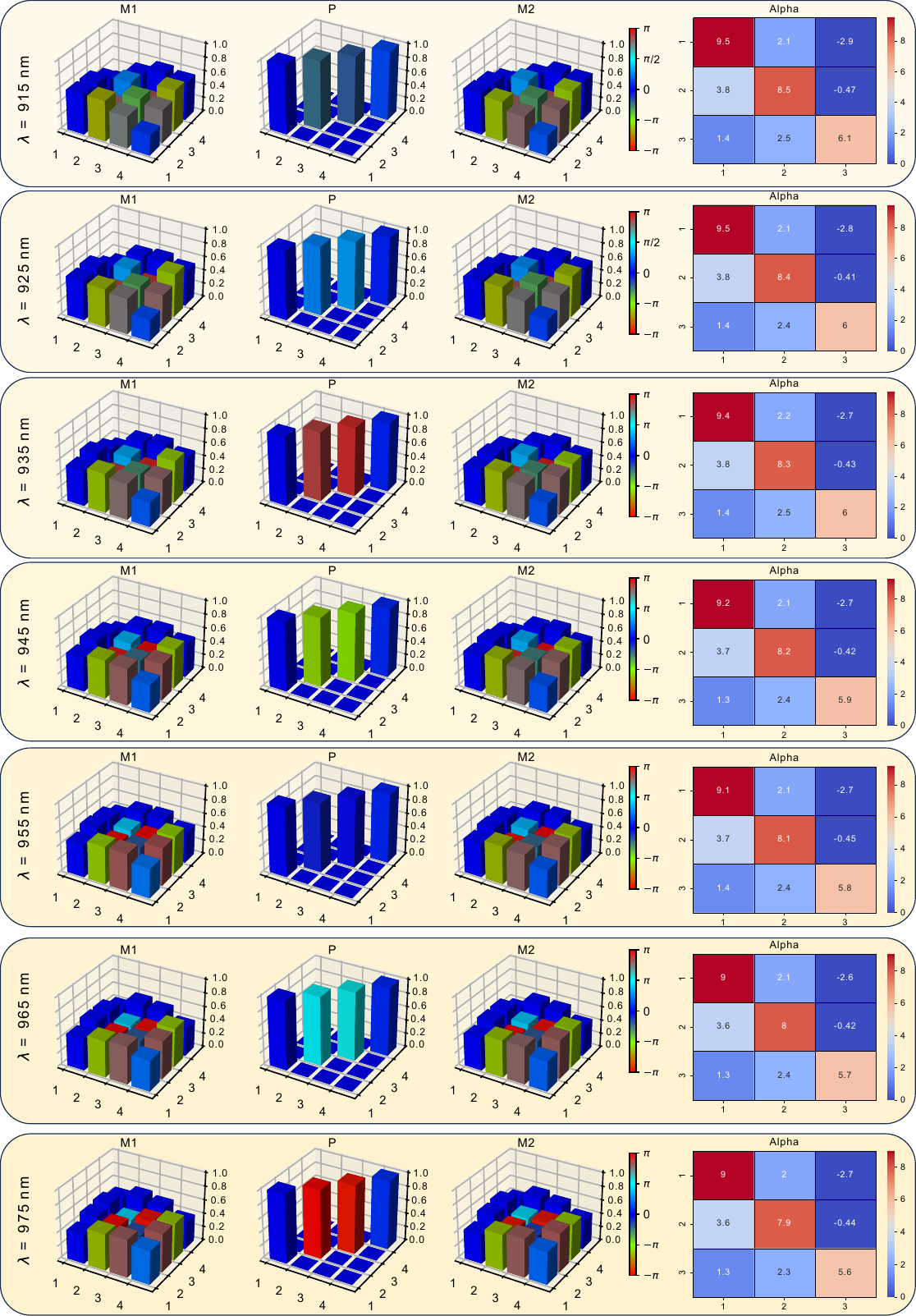}
\caption{  The reconstructed mode mixing matrices $M_{1,2}$, bias phase shift vectors $\vec{\Phi}_0$ and crosstalk matrices $A$ for each wavelength from 915 to 975 nm } 
\label{fig:all_matrices_M_P_A}
\end{figure*}

Optical switches were realized exploiting the chips' models at several wavelengths ranging from $915$ to $975$ nm. At each wavelength, the digital chip model was reconstructed using the calibration method. The reconstructed mode mixing matrices $M_{1,2}$, bias phase shift vectors $\vec{\Phi}_0$ and crosstalk matrices $A$ for each wavelength are shown in Fig. \ref{fig:all_matrices_M_P_A}.

As shown in Fig. \ref{fig:all_matrices_M_P_A} the mode mixing matrices $M^{\lambda}_{1,2}$ are slightly different for each wavelength $\lambda$ preserving the property $M^{\lambda}_{1} \approx M^{\lambda}_{2}$. The bias phase shift vectors $\vec{\Phi}_0$ vary with wavelength $\lambda$ more significantly, but always in accordance with the actual geometry of the target interferometer: $\varphi_1^{\lambda} \approx 0$ and $\varphi_2^{\lambda} \approx \varphi_3^{\lambda}$ as discussed in the Appendix \ref{app:global_fit_details}. Meanwhile, the crosstalk matrices $A^{\lambda}$ for each wavelength $\lambda$ are approximately the same. This is because the crosstalk matrix $A$ describes the physical heaters (and is completely determined by them) and not the coherent radiation.

The necessary three phase shifts for implementing a particular port-to-port optical switch were obtained by an optimization procedure run on a standard PC (Python scipy.minimize method). Once the required phase shifts were determined from the digital model of the optical chip, they were applied to the actual device. Coherent radiation of a given wavelength was injected into the specific input port of the device. The optical power distribution at the output ports was then measured using photodiodes and compared with the target values.  
The comparison between the switching matrices $U$ and $V$ was made with respect to the variation in the amplitude fidelity measure:
\begin{equation}\label{eq::amplitude_fidelity_2}
    F_{ampl}^{(2)}(U, V) = \dfrac{\sum_{i,j} |u_{ij}v_{ij}| }{ \sqrt{\sum_{i,j} |u_{ij}u_{ij}| \sum_{i,j} |v_{ij}v_{ij}| }}
\end{equation}

\begin{figure}[t]
  \centering
  \rotatebox{-90}{
    \begin{minipage}{1.3\textwidth}
      \includegraphics[width=1.\linewidth]{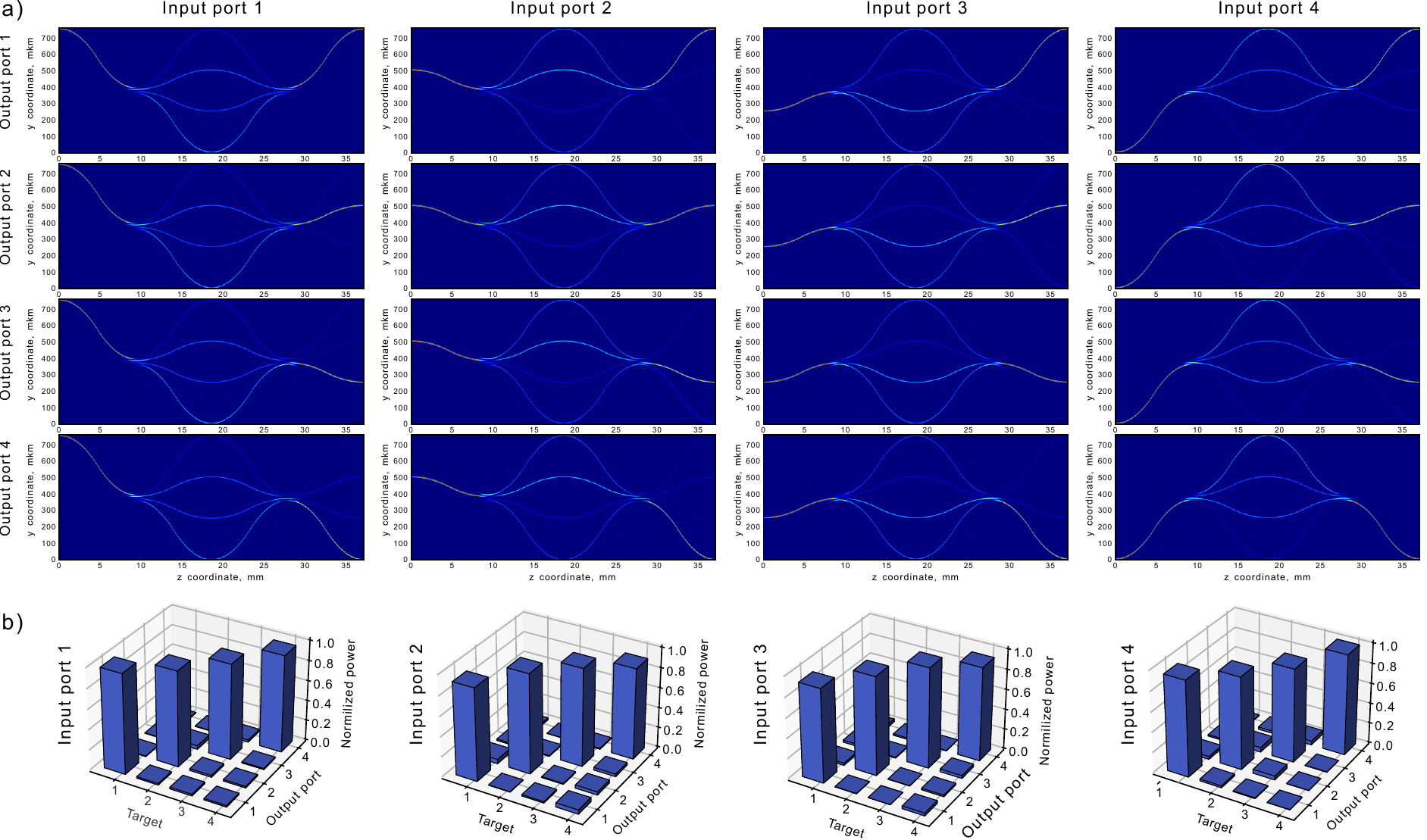}
      \caption{ a) Visualization of coherent radiation propagation through the target interferometer for realizing port-to-port optical switches for each input port to each output port of the optical chip. In each visualization a special currents set $\{ x_1, x_2, x_3 \}$ is applied to chips' tunable phase shifters in order to realize a corresponding switch. b) Histograms of the power distribution in the output ports of a device optimized for one-to-one switching to a specific output port.  }\label{fig:visualization_switches_albumn}
    \end{minipage}}
\end{figure}

An example of simulating coherent light with a wavelength of $975$ nanometers propagating through a target interferometer, illustrating all possible port-to-port optical switch configurations, is shown in Fig. \ref{fig:visualization_switches_albumn}.

\section{Scaling of the proposed calibration procedure}\label{app:scaling}

The scalability of the proposed approach for programming a reconfigurable integrated interferometer is a crucial consideration that needs to be carefully addressed.
First, in our experiment, the tunable integrated interferometer consists of two passive mode mixing blocks $M_{1,2}$ and an active phase shift layer $P_1(\vec{\varphi}_1)$ between them. The phase shift layer consists of three thermos-optical phase shifters. Therefore, our optical chip has only three tunable parameters and cannot be considered a $4\times 4$ universal photonic device.
However, if two additional phase shift layers $P_{2}(\vec{\varphi}_2)$ and $P_{3}(\vec{\varphi}_3)$ were cascaded between the two additional constant mode mixing blocks $M_3$ and $M_4$ and added to our device, the resulting interferometer would become the universal interferometer \cite{tang2017integrated, zhou2018tunable, Robust2020}.

The three-phase shift layer interferometer can be described as follows:
\begin{equation}\label{eq::universal_4x4_robust}
    U(\vec{\varphi}) = M_4 \times P_3(\vec{\varphi}_3) \times M_3 \times P_2(\vec{\varphi}_2) \times M_2 \times P_1(\vec{\varphi}_1) \times M_1,
\end{equation}
where $\vec{\varphi}_1 = \{ \varphi_1, \varphi_2, \varphi_3\}$, $\vec{\varphi}_2 = \{ \varphi_4, \varphi_5, \varphi_6\}$ and $\vec{\varphi}_3 = \{ \varphi_7, \varphi_8, \varphi_9\}$.
The procedure for programming a universal interferometer with three-phase shift layers (\ref{eq::universal_4x4_robust}) is similar to that of a single-phase shift layer device, except that a calibration routine must be performed three times for each phase shift layer $P_{1,2,3}$. Each calibration measurement will provide not only the phases of the current dependencies (including all crosstalk), but also information about each mode mixing block $M_{1,2,3,4}$.

\begin{figure}[h!]
\centering
\includegraphics[width=1\linewidth]{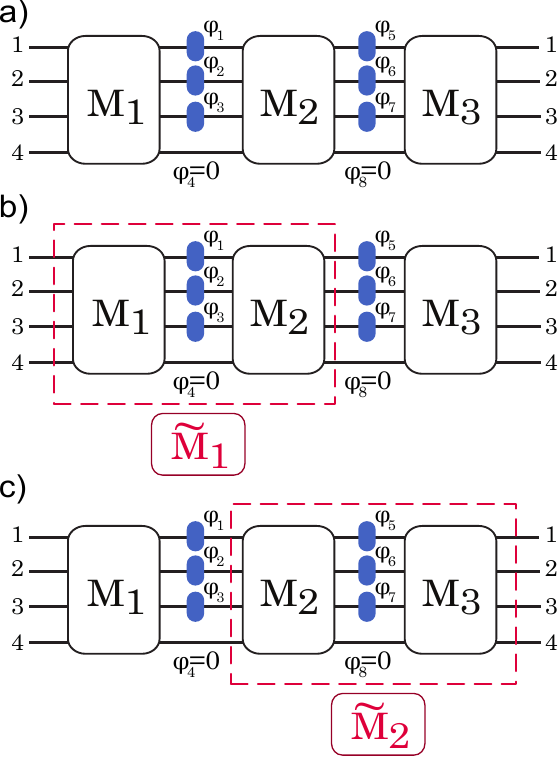}
\caption{ a) A model of an interferometer with two phase shift layers ($P_1$ and $P_2$) and three mode mixing layers ($M_1$, $M_2$ and $M_3$) can be consequently treated as b) interferometer with one phase shift layer $P_2$ cascaded between two mode mixing layers $\tilde{M}_1$ and $M_3$, and as c) interferometer with one phase shift layer $P_1$ cascaded between two mode mixing layers $M_1$ and $\tilde{M}_2$. Mode mixers, $M_1$ and $M_3$, and phase shift layers, $P_1$ and $P_2$, can be evaluated from two independent calibration routines described in the main text. The mode mixing block $M_2$ can then be implicitly determined from the auxiliary $\tilde{M}_1$ and $\tilde{M}_2$.} 
\label{fig:scaling_exmpl}
\end{figure}

To better understand how a complete digital model of an optical chip can be created from subsequent calibration measurements, let us consider an interferometer with two phase shift layers as an example (see Fig.\ref{fig:scaling_exmpl}a):
\begin{equation}\label{eq::double_4x4_robust}
    U(\vec{\varphi}) = M_3 \times P_2(\vec{\varphi}_2) \times M_2 \times P_1(\vec{\varphi}_1) \times M_1.
\end{equation}
Typically, the phase shift crosstalk between different phase shift layers is negligible, as these layers are located far enough apart from each other compared to the distance between the heaters within one phase shift layer. Therefore, each phase shift layer can be calibrated independently. The calibration of the first phase shift layer results in an optical chip model:
% \begin{equation}\label{eq::double_4x4_robust_first}
%     U(\vec{\varphi}) = \underbrace{\Tilde{M}_2}{M_3 \times P_2(\vec{\varphi}^{(0)}_2) \times M_2} \times P_1(\vec{\varphi}_1(\vec{X}_1)) \times M_1.
% \end{equation}
\begin{equation}\label{eq::double_4x4_robust_first}
    U(\vec{X}_1) =  \underbrace{M_3 \times P_2(\vec{\varphi}^{(0)}_2) \times M_2}_{\Tilde{M}_2 = \text{ const}}
    \times P_1(\vec{\varphi}_1(\vec{X}_1)) \times M_1.
\end{equation}

This calibration will result in the exact phase-current dependence for the first phase shift layer, as well as in the explicit form of the first mode mixing block $M_1$ and the effective mode mixing block $M_2$: $$\Tilde{M}_2 = M_3 \times P_2(\vec{\varphi}^{(0)}_2) \times M_2,$$ which implicitly contains the information about the actual $M_2$.
Similarly, the calibration of the second phase shift layer results in an optical chip model:
\begin{equation}\label{eq::double_4x4_robust_second}
    U(\vec{X}_2) = M_3  \times P_2(\vec{\varphi}_2(\vec{X}_2)) \times \underbrace{M_2 \times P_1(\vec{\varphi}^{(0)}_1) \times M_1}_{\Tilde{M}_1 = \text{ const}}.
\end{equation}

This calibration will result in the exact phase-current dependence for the second phase shift layer, as well as in the explicit form of the third mode mixing block $M_3$ and the effective mode mixing block $M_1$: $$\Tilde{M}_1 = M_2 \times P_1(\vec{\varphi}_1^{(0)}) \times M_1,$$ which, again, just as $\Tilde{M}_2$ implicitly contains information about the actual $M_2$.
However, we know the explicit forms of $M_1$ and $P_1(\vec{\varphi}_1^{(0)})$ (which is the first phase shift layer with zero currents). This enables us to determine the actual second mode mixing block $M_2$ as follows:
\begin{equation}\label{eq::double_4x4_robust_M_2_1}
    M_2 = \tilde{M}_1 \times M_1^{-1} \times (P_1^{(0)}) ^{-1}.
\end{equation}
Moreover, the second mode mixing block $M_2$ can also be independently obtained from $\Tilde{M}_2$:
\begin{equation}\label{eq::double_4x4_robust_M_2_2}
    M_2 = (P^{(0)}_2)^{-1} \times M_3^{-1} \times  \Tilde{M}_2.
\end{equation}
The expressions (\ref{eq::double_4x4_robust_M_2_1}) and (\ref{eq::double_4x4_robust_M_2_2}) for the calculation of $M_2$ are independent, since the entities $M_1$, $P_1^{(0)}$, $\Tilde{M}_2$ and  $M_3$, $P_2^{(0)}$, $\Tilde{M}_1$ were obtained from two independent experimental calibration procedures for the first (\ref{eq::double_4x4_robust_first}) and second (\ref{eq::double_4x4_robust_second}) phase shift layers $P_1(\vec{\varphi}_1(\vec{X}_1))$ and $P_2(\vec{\varphi}_2(\vec{X}_2))$, respectively.
The two independent evaluations of $M_2$ must be equivalent and can be used to validate the entire optical chip digital model. Figure \ref{fig:scaling_exmpl} schematically illustrates the procedure described above for programming an interferometer with two phase shift layers.

Therefore, calibrating the interferometer with a similar (complex) waveguide structure, but with $m$ phase shift layers instead of just one, would require performing $m$ separate calibration procedures for each phase shift layer as described in this paper. Such a set of measurements would be sufficient for complete characterization and programming of the optical chip.

The necessary number of measurements for the calibration of a phase shift layer in a general $N \times N$ interferometer is $N\times (N - 1)$ as a product of the input port number and the number of heaters. This scaling factor is quadratic with respect to the number of modes. 
The number of real parameters required for optimization is also quadratic in the number of modes.

In other words, there are two ways to scale an interferometer: by increasing its length, or by increasing the width or the number of modes it contains. The first method is simpler and involves a linear increase in the number of measurements and amount of processing required. However, scaling by increasing the number of modes grows quadratically with both the number of measurements and the number of parameters of the model that need to be optimized.

While the number of measurements cannot be decreased by increasing the number of modes, $N$, the number of parameters in the model can be reduced by carefully adjusting the model.
Therefore, finding possible methods to reduce the number of model parameters is of great importance.
For example, using prior knowledge of the structure of actual interferometers and the geometry of their waveguides can help reduce the number of parameters in the model.
For example, we used a physically inspired parametrization of the mode mixing matrices $M_{1,2}$ in the form of coupled waveguide lattices, which required only four real parameters per matrix, instead of nine, for the universal triangular MZI mesh parametrization. With this parameterization of $M_{1,2}$ and the same parameterization of the phase shift matrix, the total number of real parameters in our model is reduced from 30 to 20, making it significantly easier for the optimizer to find the optimal solution.

Technological advancements will also reduce the complexity of digital models.
For example, manufacturing of thermal phase shifters without mutual cross-interference eliminates all non-diagonal parameters  $\alpha_{ij}$ from the crosstalk matrix $A$, resulting in $N-1$ real parameters, which are the diagonal elements of $A$, instead of $N-1 \times N-1$  for $A$ with all nonzero elements. Alternatively, heaters can be designed to be straight and long relative to their width to simplify the model (\ref{eq::phi_with_crosstalk_matrix}) of the phase shift layer to one with only two independent parameters (\ref{eq::crosstalk_relation_appendix}), $\alpha_0$ and $\xi = d/r_0$, instead of $N-1 \times N-1$ values of $\alpha_{ij}$ (see Appendix \ref{app:global_fit_details}).

%\section{Machine learning based method for chip programming}\label{app:ML_details}

% \begin{figure*}[t!]
% \centering
% \includegraphics[width=1\linewidth]{Figs/Phases_Currents_ML.pdf}
% \caption{ The fidelity distribution histograms for ML based model test. a) Testing the learning of artificial neural network on phase shifts-matrices ($\vec{\varphi}$-$U$) dataset. b)  Testing the learning of artificial neural network on currents-matrices ($\vec{x}$ - $U$) dataset. } 
% \label{fig:ML_histograms}
% \end{figure*}

% \newpage

\end{document}